\documentclass[10pt,aps,prx,twocolumn,amsmath,amssymb,
longbibliography,
]{revtex4-2}
\usepackage{graphicx}
\usepackage{epsfig}
\usepackage{dcolumn}
\usepackage{bm}
\usepackage{bbm}
\usepackage[normalem]{ulem}
\usepackage{color}
\usepackage{dsfont}
\usepackage{float}
\usepackage[colorlinks=true,linkcolor=blue,citecolor=blue,urlcolor=blue,breaklinks=true]{hyperref}
\usepackage{verbatim}
\usepackage{subfigure}
\usepackage[ruled,vlined]{algorithm2e}
\usepackage{tikz}
\usepackage{braket}

\def\Re{{\rm Re}}

\def\be{\begin{equation}}       \def\ee{\end{equation}}
\def\bea{\begin{eqnarray}}      \def\eea{\end{eqnarray}}
\def\ba{\begin{array}}
\def\ea{\end{array}}
\def\bnum{\begin{enumerate} }
\def\enum{\end{enumerate}}

\def\=>{\Rightarrow}
\def\>{\rightarrow}

\def\eye2{Fathbb{I}}

\usepackage{ listings }

\def\Eq#1{Eq.~(\ref{#1})}
\def\Fig#1{Fig.~\ref{#1}}

\renewcommand{\v}[1]{{\bf #1}}

\renewcommand{\>}{\rangle}

\renewcommand{\Re}{{\rm Re}}

\newcommand{\eq}[2]{
	\begin{equation}
	#1 \label{#2}
	\end{equation}
}

\newcommand{\mi}{\mathrm{i}}

\renewcommand{\rm}[1]{\mathrm{#1}}

\newcommand{\vect}[1]{\boldsymbol{#1}}

\usepackage{listings}
\usepackage{color}
\definecolor{lightgray}{gray}{1}

\newcommand{\ptheta}{{\vect{\theta}}}
\newcommand{\ppsi}{{\psi_{\vect{\theta}}}}
\newcommand{\ppsistar}{{\psi^*_{\vect{\theta}}}}

\newcommand{\config}{{\vect{x}}}
\newcommand{\avg}[1]{{\left<#1\right>}}

\newcommand{\Navg}[1]{{\left[#1\right]}}
\newcommand{\supp}{{\text{supp}}}

\newcommand{\metric}{{\avg{\mathcal{I}_\lambda(\config)}_\lambda}}
\usepackage{orcidlink}

\begin{document}

\title{
Removing nodal and support-mismatch pathologies \\ 
in Variational Monte Carlo via blurred sampling
}

\author{Zhou-Quan Wan\orcidlink{0009-0007-6260-8715}}
\email{zwan@flatironinstitute.org}
\affiliation{Center for Computational Quantum Physics, Flatiron Institute, New York, NY 10010, USA}

\author{Roeland Wiersema\orcidlink{0000-0002-0839-4265}}
\email{rwiersema@flatironinstitute.org}
\thanks{Z.-Q.W. and R.W. contributed equally to this work.}
\affiliation{Center for Computational Quantum Physics, Flatiron Institute, New York, NY 10010, USA}

\author{Shiwei Zhang\orcidlink{0000-0001-9635-170X}}
\email{szhang@flatironinstitute.org}
\affiliation{Center for Computational Quantum Physics, Flatiron Institute, New York, NY 10010, USA}

\begin{abstract}
Variational Monte Carlo (VMC) is a powerful and fast-growing method for optimizing and evolving parameterized many-body wave functions, especially with modern neural-network quantum states. 
In practice, however, the stochastic estimators that form the backbone of the method can become unstable or biased due to the presence of nodes, a ubiquitous feature of quantum wave functions.
In the continuum, this results in heavy-tailed estimators with potentially divergent variances, while in discrete Hilbert spaces the sampling distribution can miss parts of the support needed to form unbiased estimators. 
These statistical pathologies lead to unreliable optimization trajectories in stochastic reconfiguration or incorrect variational dynamics in time-dependent Variational Monte Carlo (t-VMC), and severely limit the power of the numerical simulations.
We introduce blurred sampling to address these difficulties. 
The method has a number of rigorous properties that make it well-behaved, effective and computationally efficient.
Additionally, it is a post-processing approach that can be used without modifying the underlying sampler and incurs only minimal overhead.
We demonstrate its effectiveness in a broad range of problems -- including ground state optimization in both continuum and discrete settings, as well as in time-dependent variational Monte Carlo, where it resolves previously identified failure modes -- and further apply it to large-scale problems in spin dynamics.
This work establishes a broadly applicable framework for robust VMC and t-VMC calculations.
\end{abstract} 
\date{\today}
\maketitle

\begin{figure*}[htb!]
    \centering    \includegraphics[width=1.0\linewidth]{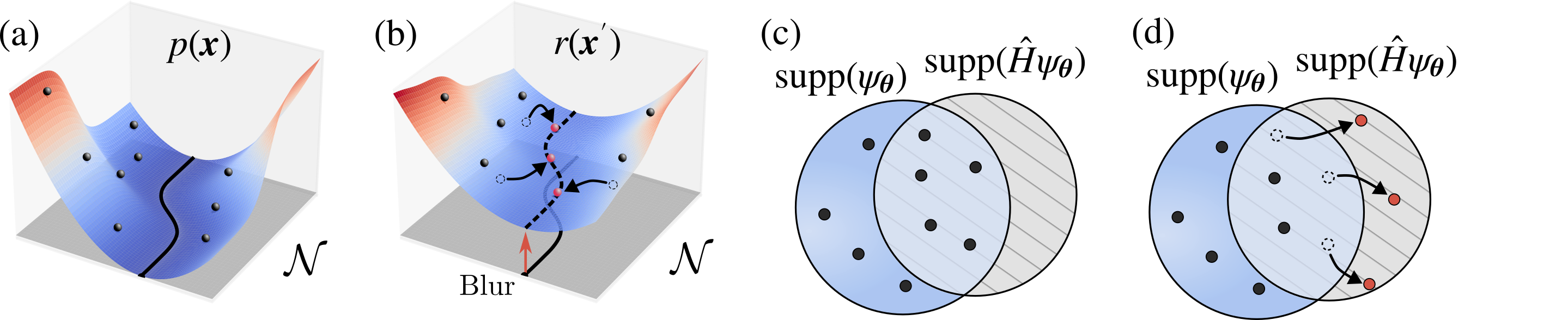}
    \caption{Statistical pathologies and their resolution via blurred sampling.
    (a) Nodal hypersurfaces lead to divergences in ratio-type estimators due to vanishing probability density $p(\config)$.
    (b) Blurred sampling locally perturbs the configurations, assigning finite probability to the original nodal set, regularizing the divergence.
    (c) The supports of $\ppsi$ and $\hat H \ppsi$ may not coincide (shaded region), resulting in a support-mismatch bias;  see \Eq{eq:bias}.
    (d) Blurred sampling exploits the connectivity of $\hat H$, allowing configurations in $\text{supp}(\ppsi)$ to access the mismatched region of $\text{supp}(\hat H \ppsi)$, eliminating the bias. 
   }
    \label{fig:Fig1}
\end{figure*}

\section{Introduction} 
Variational Monte Carlo (VMC) provides a powerful framework for studying interacting quantum many-body systems by combining variational parameterizations of the wave function with stochastic sampling, thereby circumventing the exponential growth of Hilbert space ~\cite{McMillan1965,Ceperley1980}.
Recent advances in highly expressive variational ans\"atze—most notably neural-network quantum states—together with rapid progress in machine learning techniques and computational infrastructure, have substantially expanded the scope of VMC~\cite{Carleo2017}.
Within a unified Monte Carlo framework, VMC now enables accurate calculations of ground and excited states, real- and imaginary-time dynamics, and finite-temperature properties \cite{Carleo2017,{ChooPRL2018,pathak2021excited,entwistle2023electronic,lange2024neural},{Carleo2014pra,schmitt2020quantum,Gutierrez2022realtimeevolution,Schmitt2022sciadv,schmitt2025simulating},{Sorella1998prl_sr},{Irikura2020prr,leiwangPRL2023,Nys2024prb,nys2025arxiv}}.
As a result, VMC has found broad applications across condensed matter physics~\cite{DiPRL2019,ImadaPRX2021,WenyuanPRB2021,JavierPANS2022,medvidovic2024neural,chen2024minSR,lange2024architectures,gu2025solvinghubbardmodelneural,WenyuanPRL2025,chen2025pfaffian,yixiaoArXiv2025_TI,LucianoArXiv2026}, 
quantum chemistry~\cite{hermann2020deep,PfauPRR2020,choo2020fermionic,hermann2023ab,GlehnArXiv2023,shang2024llm,PfauScience2024}, 
cold-atom or electron gas systems~\cite{kim2024neural,WilsonPRBheg2023,CassellaPRL2023,ConorPRL2024,GabrielPRB2024,Max2025prb,Yi2025prb,yubing2025prl,li2025chiralsc,conor2025disorder,conor2026arXiv} and quantum simulation~\cite{medvidovic2021classical, king2025beyond, haghshenas2025digital}.

Despite this versatility, the reliability of VMC ultimately hinges on the statistical properties of its Monte Carlo estimators.
Central quantities entering stochastic reconfiguration and time-dependent variational dynamics—such as energy gradients and variational forces—are expressed as ratios of wave-function amplitudes~\cite{Sorella1998prl_sr,Sorella2001prb,Carleo2017,schmitt2020quantum}. 
Whenever the wave function develops nodes, these ratios become singular. 
Such nodal structures are ubiquitous in fermionic systems, frustrated magnets, and systems with sign structure enforced by symmetry or gauge fields. 
Near nodes, statistical fluctuations can become heavy-tailed and may even acquire infinite variance in continuum settings~\cite{Trail2008,shihao2016,Gareth2019tailregression,wanPRE2025}. 
In discrete configuration spaces, a related but more severe pathology can arise: if the support of the wave function does not coincide with that of the Hamiltonian action, the resulting estimators remain biased even in the infinite-sample limit~\cite{Filippo2023tVMC,Carleo2022tdvp}. 
These statistical pathologies directly destabilize gradient-based optimization and variational time evolution, limiting the robustness and reliability of VMC.

Considerable effort has therefore been devoted to mitigating these instabilities. 
Some approaches modify the estimator directly through explicit regularization~\cite{Wagner2020regularization} or variance-reduction techniques such as control variates~\cite{Filippo2023tVMC,gravina2025neural,linteau2026variancereduction}. 
More commonly, the problem is addressed through importance-sampling strategies that replace the original sampling distribution by a modified reference measure~\cite{eric2017variancematching,Eric2020excitedstate,InuiPRRalphasampling,misery2025lookingelsewhereimprovingvariational,ASmethod,Trail2008alternativesampling}. 
While these approaches can alleviate specific pathologies, they introduce competing requirements
which are difficult to satisfy simultaneously.
In practice, residual bias or variance amplification persists, and significant modification to the algorithm and additional computational overhead are often incurred. 
A broadly applicable and structurally robust solution to nodal instabilities in VMC remains lacking, which hinders progress in many application areas.

In this work, we introduce \emph{blurred sampling} to address this challenge. 
Figure~\ref{fig:Fig1} provides a schematic illustration of the new approach.
Rather than adding explicit regularization to the estimator or globally redefining the sampling distribution, blurred sampling applies a single local mixing step to Monte Carlo configurations as a post-processing operation. 
This procedure induces an implicit reference measure that regularizes nodal singularities at their origin while leaving the underlying sampling dynamics unchanged. 
We show that the resulting reweighting factors are strictly bounded, which ensures a finite effective sample size and resolves both infinite-variance behavior and support-mismatch bias. 
Because the method is implemented as a post-processing step, it preserves the computational scaling of standard VMC and, in discrete configuration spaces, can often be realized without additional wave-function evaluations. 
This design allows blurred sampling to integrate seamlessly with existing variational ansätze, optimization schemes, real- and imaginary-time dynamics, and modern sampling architectures~\cite{malyshev2024neuralquantumstatespeaked,liu2025efficientoptimizationneuralnetwork}, including autoregressive models~\cite{uria2016neural,hibatallah2020,Sharir2020prl,Stephan2023scipost,sprague2024variational,moss2025leverage1, moss2025leverage2}.

As we demonstrate below through theoretical analysis and numerical examples, 
blurred sampling restores stable and accurate simulations in regimes where conventional VMC breaks down. It provides a broadly applicable and scalable framework for achieving reliable simulations in variational Monte Carlo and neural quantum states, which will enable a wide range of applications across disciplines in quantum science.

\section{Background and Preliminaries}

\subsection{Variational Monte Carlo}\label{subsec:VMC}

In the general VMC framework, the many-body state is approximated by a parameterized wave function $\ket{\ppsi}=\sum_{\config}\ppsi(\config)\ket{\config}$ in the computational basis $\{\ket{\config}\}$, where the variational parameters $\boldsymbol{\theta}\in \mathbb{C}^{N_\text{para}}$ are taken to be complex-valued, with real parameterizations recovered as a special case, and the sum over $\config$ can be either discrete or continuous (integral).
The optimization of the variational energy $ E_{\boldsymbol{\theta}}\equiv \frac{\braket{\ppsi|\hat H|\ppsi}}{\braket{\ppsi|\ppsi}}$ and the real- or imaginary-time dynamics are governed by two fundamental objects: the variational force $\boldsymbol{F}\equiv \partial_{\boldsymbol{\theta^*}}E_{\boldsymbol{\theta}}$ and the quantum geometric tensor (QGT) $\boldsymbol{S}$. 
They are defined as
\eq{
\begin{split}
F_i &= \frac{\braket{\partial_{\theta_i}\ppsi|\hat H|\ppsi}}{\braket{\ppsi|\ppsi}}-\frac{\braket{\partial_{\theta_i}\ppsi|\ppsi}}{\braket{\ppsi|\ppsi}}\frac{\braket{\ppsi|\hat H|\ppsi}}{\braket{\ppsi|\ppsi}},\\
S_{ij} &= \frac{\braket{\partial_{\theta_i}\ppsi|\partial_{\theta_j}\ppsi}}{\braket{\ppsi|\ppsi}}-\frac{\braket{\partial_{\theta_i}\ppsi|\ppsi}}{\braket{\ppsi|\ppsi}}\frac{\braket{\ppsi|\partial_{\theta_j}\ppsi}}{\braket{\ppsi|\ppsi}}.
\end{split}
}{eq:force_and_QGT}
These quantities determine the dynamics of parameters in stochastic reconfiguration (SR) and time-dependent VMC (t-VMC) through
\eq{
S_{ij}\dot{\theta}_j = -\xi F_i,
}{eq:tdvp_equation}
which is known as the time-dependent variational principle (TDVP) equation~\cite{Sorella1998prl_sr, carleo2012localization, Carleo2017, schmitt2020quantum, schmitt2025simulating}. The choice $\xi=1$ corresponds to the imaginary-time evolution (i.e., SR), and $\xi=i$ yields real-time dynamics.

In VMC, these quantities are evaluated by Monte Carlo sampling from the Born distribution $p(\config)=|\psi_{\ptheta}(\config)|^2/\sum_\config|\psi_{\ptheta}(\config)|^2 $.
The estimators take the covariance form
\eq{
\begin{split}
F_i^\text{MC} &= \avg{\mathcal{O}_i^*(\config) E_\text{loc}(\config)}_{p}-\avg{\mathcal{O}^*_i(\config)}_{p}\avg{E_\text{loc}(\config)}_{p},\\
S_{ij}^\text{MC} &= \avg{\mathcal{O}_i^*(\config) \mathcal{O}_j(\config)}_{p}-\avg{\mathcal{O}^*_i(\config)}_{p}\avg{\mathcal{O}_j(\config)}_{p},
\end{split}
}{eq:mc_estimator}
where $\avg{\cdot}_p $ denotes expectation with respect to $p(\config)$, $\mathcal{O}_i(\config)\equiv
\partial_{\theta_i}\psi_{\ptheta}(\config)/\psi_{\ptheta}(\config)$ is the logarithmic derivative, and $E_\text{loc}(\config)\equiv \braket{\config|\hat H|\ppsi}/\ppsi(\config)$ is the local energy. 

This representation makes the origin of statistical pathologies transparent: both $E_\text{loc}$ and $\mathcal{O}_i$ contain divisions by $\ppsi(\config)$, so the corresponding ratio-type estimators can diverge in the vicinity of wave-function nodes, i.e., when $|\ppsi(\config)|\to 0$.
As a consequence, the stochastic estimators entering VMC can often become ill-behaved, as we discuss further in the next section.

\begin{figure}[t]
    \centering    \includegraphics[width=1.0\columnwidth]{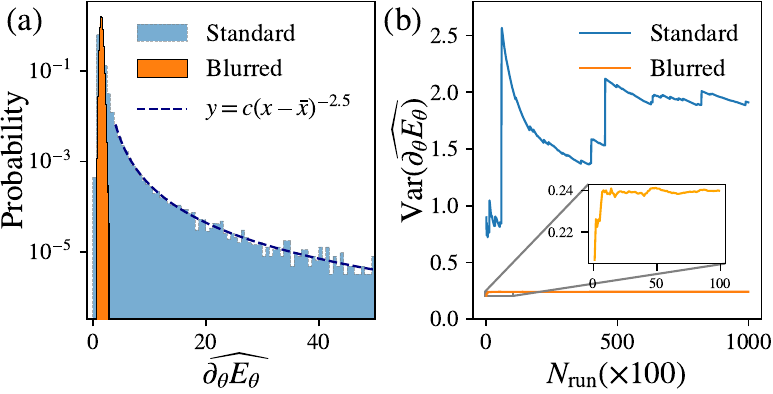}
    \caption{Pedagogical example in the continuum. 
    We consider two non-interacting spinless fermions on a ring, with the Hamiltonian $\hat H = -\frac{1}{2}(\partial^2_1+\partial^2_2) $, and variational ansatz $\ppsi(x_1, x_2)=\cos(\theta) \sin(x_1-x_2) + \sin(\theta) \sin(2x_1-2x_2)$, $x_{1,2}\in[0,2\pi]$. 
    (a) Distribution of the gradient estimator based on 1000 samples at $\theta=\pi/4$.  Standard sampling exhibits a heavy-tailed distribution with exponent $\alpha = 1.5$, while blurred sampling yields a finite-variance estimator. 
    (b) Estimated variance versus number of runs; Standard sampling does not converge.
   }
    \label{fig:Fig2}
\end{figure}
\subsection{Statistical Pathologies}
\label{ssec:pathologies}
These pathologies take different forms depending on the structure of the configuration space.
We distinguish between continuum and discrete settings in turn.

\paragraph*{Infinite variance in continuum systems.} \label{subsec:infinite_variance}
In continuum fermionic systems, the wave function $\ppsi(\config)$ depends continuously on the particle coordinates.
Fermionic statistics enforces the presence of nodal sets
$\mathcal{N}\equiv\{\config| \ppsi(\config)  =0\}$ (see \Fig{fig:Fig1}(a)), whose geometry has been extensively studied
\cite{ceperley1991fermion,foulkes2001quantum,Mitas2006prl,needs2009continuum}.
In the vicinity of a node, the wave function vanishes linearly in the normal direction, $|\ppsi|\propto d$, where $d$ denotes the distance to the nodal set.
Consequently, ratio-type quantities in \Eq{eq:mc_estimator} diverge as $|E_\text{loc}|,\,|\mathcal{O}_i| \propto 1/d$.
Since the distribution scales as $p(\config) \propto d^2$ near the node, the resulting estimator can have finite expectation but infinite variance.

To illustrate the resulting pathology, we consider a minimal continuum fermionic system (two spinless fermions on a ring).
As shown in Fig.~\ref{fig:Fig2}(a), the distribution of the energy-gradient estimator exhibits a pronounced heavy tail~\cite{levy1925calcul,gnedenko1968limit, nolan2020univariate,Antoine1990anomalous}.
More precisely, the tail follows $P(|X|>x)\sim C x^{-\alpha}$ as $x\to\infty$, with exponent $\alpha\simeq1.5$.
\Fig{fig:Fig2}(b) shows that the variance fails to converge with increasing Monte Carlo runs.
This behavior is generic. For real (complex) wave functions, the tail exponent of the estimator in \Eq{eq:mc_estimator} satisfies $\alpha=3/2$ ($\alpha=2$), reflecting the codimension of the nodal manifold~\cite{Trail2008}.
Since $\alpha\le 2$, the variance diverges, leading to anomalously slow and unstable Monte Carlo convergence—the so-called infinite variance problem~\cite{Trail2008,shihao2016,Gareth2019tailregression,wanPRE2025} , i.e. $\epsilon \propto {1}/{N^{1/3}}$ for $\alpha = 3/2$ and $\epsilon \propto \sqrt{{\log (N)}/{N}}$ for $\alpha=2$ rather than $1/\sqrt{  N}$~\cite{nolan2020univariate}.

\begin{figure}[t]
    \centering    \includegraphics[width=1.0\columnwidth]{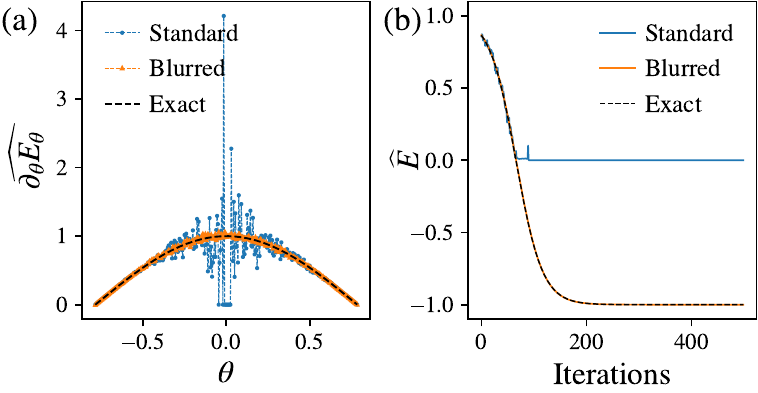}
    \caption{Pedagogical example in discrete space. 
    We consider a single-spin system with Hamiltonian $\hat H =  X$ and $\ppsi=(\cos(\theta), \sin(\theta))$. 
    (a) Gradient estimator at different $\theta$ using 1000 samples. As $\theta \to 0$, standard sampling exhibits large fluctuations and develops a systematic bias, whereas blurred sampling remains stable.
    (b) Energy optimization via stochastic reconfiguration (SR), initialized at $\theta=\pi/3$. Standard sampling can become trapped near $\theta\to 0 $, while blurred sampling follows the exact imaginary-time trajectory.
   }
    \label{fig:Fig3}
\end{figure}

\paragraph*{Bias in discrete configuration spaces.} 
In discrete configuration spaces, the problem can become more severe.
The support of $\ppsi(\config)$ need not coincide with that of $\braket{\config|\hat H|\ppsi}$, see \Fig{fig:Fig1}(c).
Configurations that contribute to the exact force may therefore lie outside the sampled support $\mathrm{supp}(\ppsi)=\{\config\,|\,\ppsi(\config)\neq0\}$.
As a result, the Monte Carlo estimator can remain biased even in the infinite-sample limit \cite{Filippo2023tVMC,Carleo2022tdvp}.
Explicitly,
\eq{
F_i-F_i^\text{MC}=\frac{\sum_{\config\notin \text{supp}(\ppsi)} \braket{\partial_{\theta_i} \ppsi|\config}\braket{\config|\hat H|\ppsi}}{\sum_{\config} |\ppsi(\config)|^2}.
}{eq:bias}
The right-hand side corresponds precisely to the contribution of configurations that are never sampled under $p$. 
We refer to this structural discrepancy as a \emph{support-mismatch bias}. Even absent an absolute mismatch, when $\ppsi(\mathbf{x})\ll \braket{\config|\hat H|\ppsi}$ on rarely sampled configurations, the ratio-type structure of the estimator can strongly amplify statistical fluctuations and render the simulation unstable.
The effect of the support-mismatch bias is illustrated in \Fig{fig:Fig3}, using a single-spin example with $\hat H=X$, and variational ansatz $\ppsi=(\cos\theta, \sin\theta)$.
As $\theta\to0$, the gradient estimator exhibits increasingly large fluctuations and eventually develops a systematic bias. 
As a result, a simple ground state optimization can lead to an incorrect answer.
These issues are not specific to the force: the QGT, sharing the same ratio-type structure, is subject to an analogous bias and instabilities.

\begin{table*}[t!]
    \centering
    \begin{tabular}{|c|c|c|c|c|c|}\hline
          Method&  Infinite variance&  Force bias&  ESS scaling&  Sampling&  Additional complexity\\\hline
         Overdispersed sampling $|\ppsi|^\alpha$~\cite{InuiPRRalphasampling,misery2025lookingelsewhereimprovingvariational}&  solved &  unsolved&  $O(e^{-N})$&  modified&  0 \\\hline
 $E_\text{loc}$ regularization~\cite{Trail2008alternativesampling}& solved& unsolved& $O(1)$& modified& $E_\text{loc}$\\\hline
 Node distance regularization~\cite{ASmethod}& solved& N/A& $O(1)$& modified& $\nabla \ppsi$\\ \hline
 Blurred sampling (this work)& solved& solved& $\geq 1-q$& postprocessing& 0; $N_\text{conn}$\\\hline
    \end{tabular}
    \caption{Comparison between different importance sampling methods and the blurred sampling approach.
    The ESS scaling refers to how the effective sample size, defined in \Eq{eq:ESS}, scales with the system size $N$. The additional complexity quantifies the extra computational cost beyond standard sampling, measured in terms of additional wave function evaluations, which typically constitute the main computational bottleneck of the algorithm. For the middle two methods, this cost is difficult to express purely in terms of wave function evaluations; instead, we list the primary additional quantity required by each method. In discrete space, blurred sampling introduces zero overhead, whereas in the continuum the cost scales as $N_\text{conn}$. 
    }
    \label{tab:comparison}
\end{table*}

\subsection{Requirements for a Remedy}
\label{ssec:requirements}

The infinite variance problem and support-mismatch bias
identified above present a number of challenges, compounded by the fact that VMC is typically applied in extremely high dimensions. A natural strategy for remedying nodes is to modify the sampling distribution via importance sampling~\cite{Rubinstein_1981}.
By drawing samples $\config'$ from a reference distribution $ r(\config')$, the variational force in \Eq{eq:force_and_QGT} can be evaluated in a self-normalized importance sampling (SNIS) form:
\eq{
F_i^\text{SNIS}=\frac{\avg{\omega\mathcal{O}_i^*E_\text{loc}}_{r}}{\avg{\omega}_{r}}-\frac{\avg{\omega\mathcal{O}_i^*}_{r}}{\avg{\omega}_{r}}\frac{\avg{\omega E_\text{loc}}_{r}}{\avg{\omega}_{r}},
}{eq:snis_estimator}
where $\omega(\config)\equiv p(\config)/r(\config)$ is the \emph{reweighting factor}.
Note that divergences arising from rarely sampled configurations
can now be compensated by the reweighting factor $\omega$.
In principle, infinite variance can be mitigated if
$r$ reduces the effective codimension of nodal set.
Similarly, the support-mismatch bias in the discrete case can be resolved provided that the reference distribution satisfies $\text{supp}(\hat H\ppsi) \cup  \text{supp}(\ppsi)\subseteq \text{supp}(r)$ for the forces and $\text{supp}(\partial_{\theta_i}\ppsi(\config)\partial_{\theta_j}\ppsi(\config))\cup \supp(\ppsi)\subseteq \text{supp}(r)$ for the QGT. 
In both cases, the missing contributions are restored.

However, identifying a proper choice of $r(\config)$ that is easy to sample,
removes the nodes, and does not significantly modify the distribution elsewhere is very difficult.
Because quantum problems typically involve extremely high-dimensional configuration spaces,
$p(\config)$ can vary by many tens of orders of magnitude across configurations~\cite{shihao2016}.
This difficulty is further compounded by the fact that modifying the sampling distribution introduces a reweighting factor
$\omega = p/r$, whose fluctuations strongly affect the statistical efficiency of the estimator.

Existing approaches construct $r$ in different ways.
For example, overdispersed sampling schemes employ
$r(\config)\propto |\psi(\config)|^\alpha$ with $\alpha<2$,
which broadens the distribution and is straightforward to sample \cite{InuiPRRalphasampling,misery2025lookingelsewhereimprovingvariational}.
However, the resulting reweighting factor
$\omega \propto |\psi|^{2-\alpha}$
exhibits increasingly large fluctuations as the system size grows,
causing an exponential degradation of the effective sample size ($\mathrm{ESS}\equiv{\avg{\omega}_r^2}/{\avg{\omega^2}_r}$)
and preventing reliable scaling to larger systems
(A quantitative example is given in App.~\ref{sec:overdispersed}).
Other methods introduce explicit regularization factors based on
local-energy estimators or nodal-distance diagnostics~\cite{Trail2008alternativesampling, ASmethod},
leading to modified distributions of the form
$r(\config)=g(\config)p(\config)$,
where $g(\config)$ is constructed to diverge near the nodal set
so as to compensate for the vanishing of $p(\config)$.
While such constructions can partially tame divergences,
they typically rely on continuum-space diagnostics
and substantially alter the target distribution.
This increases algorithmic complexity,
requires additional evaluations per configuration,
and makes the overall computational cost difficult to control,
particularly in the context of modern neural-network quantum states (NQS).

These examples illustrate that designing a suitable reference distribution
involves balancing several competing objectives.
An admissible reference distribution must simultaneously satisfy
the following requirements:
\begin{enumerate}
\setlength{\itemsep}{2pt}
\setlength{\parskip}{0pt}
\setlength{\parsep}{0pt}
\item[(i)] it should have sufficiently broad support to regularize nodal singularities;
\item[(ii)] it should control the fluctuations of $\omega(\config)$ to prevent effective-sample-size collapse;
\item[(iii)] it should remain computationally efficient for sampling and evaluation of the reweighting factor.
\end{enumerate}
These requirements are generally in tension with one another, and no existing reference distribution satisfies all of them simultaneously in a generic VMC setting (see Table~\ref{tab:comparison}).

\section{Blurred sampling method}

Overcoming these challenges requires a different strategy.
Conceptually inspired by the bridge-link method in determinant quantum Monte Carlo~\cite{shihao2016,wanPRE2025}, we introduce a local mixing scheme applied to sampled configurations as a post-processing step.
This construction, which we term \emph{blurred sampling}, modifies configurations without redefining the underlying sampling distribution. 
Intuitively, it “blurs’’ sampled configurations, replacing the singular sampling measure with a locally mixed reference distribution that assigns finite weight near nodal regions.
In effect, this produces an effective $r(\config)$ that is coupled to $p(\config)$ locally, allowing the reweighting factor $\omega(\config)$ to remain well behaved as sampling traverses the widely varying landscape of $p(\config)$.

\subsection{Construction}
Starting from a configuration $\config\sim p(\config)$, a new configuration $\config'$ is generated according to a \emph{blur kernel}, defined by the discrete Markov transition kernel 
\eq{
K(\config'|\config) = (1-q)\,\delta_{\config',\config} + q \, K_\mathrm{off} (\config'|\config),
}{eq:general_kernel}
where $ 0 \le q <1 $ controls the strength of the blurring.
Operationally, with probability $1-q$ the configuration $\config$ is left unchanged, while with probability $q$ it is updated according to the normalized, off-diagonal kernel $K_{\mathrm{off}}(\config'|\config)$, which satisfies $K_\mathrm{off}(\config|\config)=0$ and $\sum_{\config'}K_{\mathrm{off}}(\config'|\config)=1$ for any $\config$.
This post-processing step induces an implicit reference distribution for $\config'$,
\eq{
r(\config') = \sum_{\config}K(\config'|\config)p(\config).
}{eq:reference_dist}
Because all contributions to $r(\config')$ are nonnegative, we have $r(\config')>0$ whenever there exists a configuration $\config$ with $K(\config'|\config)>0$ and $p(\config)>0$. 
The blurring operation therefore expands the support of the sampling distribution according to the connectivity structure of $K_{\mathrm{off}}(\config'|\config)$, satisfying requirement (i) of Sec.~\ref{ssec:requirements}, given an appropriately chosen kernel $K(\config'|\config)$. 
The blur kernel can be chosen in a variety of ways; below we present representative constructions for continuum and discrete systems and illustrate their effectiveness using the two pedagogical examples introduced in Sec.~\ref{ssec:pathologies}.

\paragraph*{Continuum systems.}
In continuum space where $\config = \{\vec{r}_1,\cdots , \vec{r}_{n_e}\}$, singular behavior arises from the vanishing of the wave function along nodal manifolds. 
Locally, the wave function varies linearly along normal directions to the nodal set. 
It therefore suffices to introduce small local perturbations of configurations so as to assign nonzero weight to configurations in the nodal set.
To cover all local coordinate directions, one simple way is to define the connected set
\eq{
\mathcal{C}^\text{out}(\config)=\left\{
\config \pm \epsilon\,\mathbf e_{i\alpha}| i=1,\ldots,n_e,\alpha=1,\ldots,d
\right\},}{eq:continuous_kernel}
where $\mathbf e_{i\alpha}$ shifts the $\alpha$-th coordinate of particle $i$ by one unit and $\epsilon$ is a small parameter to control the noise strength.  
We then use the blur kernel 
\begin{equation}
K(\config'|\config) =
(1-q)\,\delta_{\config',\config} +
\frac{q}{N_{\mathrm{conn}}}
\sum_{\boldsymbol{y} \in \mathcal{C}^\text{out}(\config)} \delta_{\config',\boldsymbol{y}},
\label{eq:blur_kernel}
\end{equation}
where $N_{\mathrm{conn}}=|\mathcal{C}^\text{out}(\config)| =2dn_e$.
This means that for each sampled configuration $\config$, with probability $q$ one particle is randomly displaced by $\epsilon$ along one of the coordinate directions.
The connectivity therefore remains sparse and scales linearly with system size.
Since all $d\,n_e$ local coordinate directions are covered, the procedure effectively smears nodal singularities (see \Fig{fig:Fig1}(b)).
For the example shown in \Fig{fig:Fig2}, blurred sampling ($q=2/3$, $\epsilon = 0.2$) straightforwardly eliminates the infinite-variance problem.

\paragraph*{Discrete configuration space.}
In discrete spaces, the bias of the force $\boldsymbol{F}$ originates from the support mismatch between $\ppsi$ and $\hat H\ppsi$.
We therefore exploit the connectivity structure of the Hamiltonian to define the blur kernel.
Specifically, we denote by 
\eq{\mathcal{C}^\text{out}(\config)= \{\boldsymbol{y}\neq \config\mid\braket{\boldsymbol{y}|\hat H|\config}\neq 0\}}{} 
the set of configurations connected to $\config$ by $\hat H$, and adopt the same kernel as in \Eq{eq:blur_kernel}.
More refined kernels are also possible; for example,
the transition probabilities can be weighted according to the matrix elements of $\hat H$, since many Monte Carlo quantities, such as the force and energy variance, involve the same matrix elements through $E_\text{loc}$.

With this construction, the support-mismatch bias is eliminated for the variational force $\boldsymbol{F}$.
Since $\mathrm{supp}(\ppsi)\cup \mathrm{supp}(\hat H\ppsi)\subseteq \mathrm{supp}(r)$,
all configurations contributing to the force are sampled, and the resulting estimator becomes unbiased.
In the example of Fig.~\ref{fig:Fig3}, blurred sampling ($q=0.5$) removes both the bias and the large fluctuations of the gradient estimator, leading to stable and smooth ground-state optimization.

\subsection{Statistical Guarantees and Computational Scaling}
The blurred sampling method possesses two key statistical properties.
First, the reweighting factor satisfies $\avg{\omega(\config')}_{r}=1$ as a direct consequence of the Markov property  (see App.~\ref{app:unit}).
As a result, the SNIS estimator in \Eq{eq:snis_estimator} has a standard importance-sampling form, and the following estimator (an analogous expression holds for QGT)
\eq{
\widehat{F}_i^\text{blur}= \frac{N\Navg{\omega}_r^2}{N-1} \times
\left(\frac{\Navg{\omega \mathcal{O}_i^* E_\text{loc}}_r}{\Navg{\omega}_r} -\frac{\Navg{\omega \mathcal{O}_i^* }_r}{\Navg{\omega}_r}\frac{\Navg{\omega E_\text{loc}}_r}{\Navg{\omega}_r}\right)
}{eq:unbiased_estimator} 
is finite-sample unbiased, i.e. $\mathbb{E}[\widehat{F}_i^\text{blur}]=\lim_{N\to\infty}\widehat{F}_i^\text{blur}$ (see App.~\ref{app:finite_sample_unbiased}).
Here $\Navg{\cdot}_r = \frac{1}{N}\sum_{\config'} \cdot(\config')$
denotes the finite-sample average over samples drawn from $r$.
This estimator differs from the SNIS estimator only by an overall factor $\frac{N\Navg{\omega}_r^2}{N-1}$. 
Consequently, in solving the TDVP equation in \Eq{eq:tdvp_equation}, this common factor cancels in the linear system.
Blurred sampling can therefore be implemented operationally
in a self-normalized form while retaining finite-sample unbiasedness.

Second, the fluctuations of $\omega(\config')$ are explicitly controlled.
From the construction of the blur kernel in \Eq{eq:general_kernel}, the reweighting factor is bounded as $0\leq\omega (\config')\leq 1/(1-q)$.
This boundedness immediately implies a lower bound on the effective sample size (ESS) (see App.~\ref{app:bound}):
\eq{
\mathrm{ESS}(\omega(\config'))\equiv
\frac{\avg{\omega(\config')}_r^2}{\avg{\omega(\config')^2}_r} \geq 1-q.
}{eq:ESS}
Thus, for any fixed $0<q<1$, the effective sample size remains finite, 
preventing its exponential degradation with increasing system size.
Consequently, the blurred sampling method avoids the finite-sample bias inherent to SNIS, and satisfies requirement (ii) outlined in Sec.~\ref{ssec:requirements}.
Importantly, the method is robust to the precise choice of $q$: for a broad range of $q\sim O(1)$, the estimator remains stable, as will be demonstrated in the numerical results later.

The blurred sampling method also exhibits well-controlled computational cost.
The blurring step is a \emph{post-processing} procedure and does not require additional wave-function evaluations; hence, it does not inherit the computational complexity of the underlying ans\"atze and can therefore be combined seamlessly with sophisticated sampling methods \cite{malyshev2024neuralquantumstatespeaked,liu2025efficientoptimizationneuralnetwork} and autoregressive networks~\cite{uria2016neural,hibatallah2020,Sharir2020prl,Stephan2023scipost,sprague2024variational,moss2025leverage1, moss2025leverage2}.
The only additional cost, measured in units of wave-function evaluations (the dominant expense in VMC), arises from computing the reweighting factor
\eq{
\omega(\config')= \frac{p(\config')}{(1-q)p(\config')+\sum_{\config \in \mathcal{C}^\text{in}(\config')} K(\config'|\config) p(\config) },
}{eq:reweighting_factor}
where $\mathcal{C}^\text{in}(\config')=\{\config|\, K(\config'|\config)>0\}$ denotes the set of configurations connected to $\config'$ through the kernel.
Evaluating $\omega(\config')$ therefore requires $|\mathcal{C}^\text{in}(\config')|$ wave-function evaluations.
To preserve computational scalability, we restrict the blur kernel to be sparsely connected, such that $|\mathcal{C}^\text{in}(\config')|$ grows at most polynomially with system size.
Under this constraint, blurred sampling does not alter the overall computational scaling of standard VMC, thereby satisfying requirement (iii) in Sec.~\ref{ssec:requirements}.

In the continuum case, the kernel introduced above is symmetric and sparsely connected, with 
$|\mathcal{C}^{\text{in}}(\config)| = |\mathcal{C}^{\text{out}}(\config)| = 2 d n_e$, 
which scales linearly with system size.
In discrete configuration spaces, evaluating the reweighting factor in Eq.~\eqref{eq:reweighting_factor} requires amplitudes of the wave-function only on configurations connected by $\hat H$. 
These amplitudes are already computed in the evaluation of the local energy. 
Consequently, blurred sampling in the discrete case introduces essentially no additional computational cost.

\subsection{Generalizations}\label{sec:randomized_blur}
The constructions introduced above rely on a fixed blur kernel with a prescribed connectivity structure. 
While this deterministic choice suffices to eliminate the dominant pathologies—such as infinite variance and bias in the variational force—there remain situations in which a broader exploration of configuration space may be desirable.
One possibility would be to employ more densely connected kernels, with the associated reweighting factor estimated through an auxiliary Monte Carlo procedure \cite{shihao2016}. 
Although such constructions expand the accessible support, they involve an additional stochastic layer (Alternative to the separate Monte Carlo integration, a  different computational structure~\cite{Alexandru2023prd} has been explored).

This motivates a randomized generalization of blurred sampling that enlarges the effective support while preserving computational efficiency.
Specifically, consider a family of sparse blur kernels $\{K_\lambda\}$ with a connectivity structure that is parameterized by $\lambda$, each satisfying the structure of Eq.~\eqref{eq:general_kernel}. 
For each sample $\config_i \sim p(\config)$, we independently draw $\lambda_i \sim f(\lambda)$ from a normalized density $f$, and then generate $\config'_i \sim K_{\lambda_i}(\cdot|\config_i)$.
Let $r_\lambda(\config')=\sum_{\config}p(\config)K_\lambda(\config'|\config)$ denote the blurred distribution corresponding to a fixed $\lambda$.
We define the conditional reweighting factor
$\omega_\lambda(\config')=\frac{p(\config')}{r_\lambda(\config')}$.

The estimator in Eq.~\eqref{eq:unbiased_estimator} then applies analogously under randomized blurred sampling, with $\Navg{\omega \cdot}_{r}$ replaced by $\Navg{\omega_\lambda \cdot}_{f(\lambda) r_\lambda}$. 
Through the stochastic choice of $K_\lambda$, the sampling effectively explores a broader support while preserving finite-sample unbiasedness.
Importantly, the weight is evaluated conditionally with respect to the realized kernel $K_\lambda$, so the computational cost is controlled by the sparsity of each individual kernel rather than by the averaged kernel. 
This randomized construction increases flexibility while preserving scalability.
As illustrated in later examples, such flexibility can further mitigate instabilities in the QGT and enable more refined blur kernels like a Gaussian kernel in the continuum case. 
A detailed derivation of the randomized blurred sampling scheme is provided in Appendix~\ref{app:randomized_blur} and \ref{app:qgt_blur}.

In summary, blurred sampling provides a structurally robust 
solution under the general importance-sampling scheme. 
Unlike overdispersed or explicitly regularized reference distributions (see Table~\ref{tab:comparison}), it preserves the underlying sampling dynamics and avoids reweighting factors with system-size-dependent variance growth. 
Instead, it operates as a sparse post-processing step that restores missing support and regularizes singular estimators while maintaining computational scalability.

\section{Results}\label{sec:gs_examples}

Having established the conceptual foundation of blurred sampling and validated it on pedagogical examples, we now turn to realistic settings. We first consider two ground state optimization examples in Sec.~\ref{ssec:gs-vmc}: the two-dimensional electron gas in continuum space and the $J_1$-$J_2$ model in discrete space. 
We show that blurred sampling removes the infinite-variance problem and the statistical pathologies in the Monte Carlo estimators, makes the optimization more stable, and allows faster convergence to the ground state. 
We then move on to time-dependent variational Monte Carlo (t-VMC), where statistical pathologies have been explicitly identified in the literature~\cite{Filippo2023tVMC,vrcan2025instability}.
In this setting, the correctness of the time evolution hinges directly on the unbiasedness and stability of the Monte Carlo estimators entering the TDVP equation.
We show that blurred sampling resolves previously identified failure cases and scales efficiently to large-scale spin-dynamics simulations.
Thus blurred sampling directly addresses a fundamental limitation in applications of t-VMC for quantum dynamics.

\begin{figure*}
    \centering
    \includegraphics[width=1.0\linewidth]{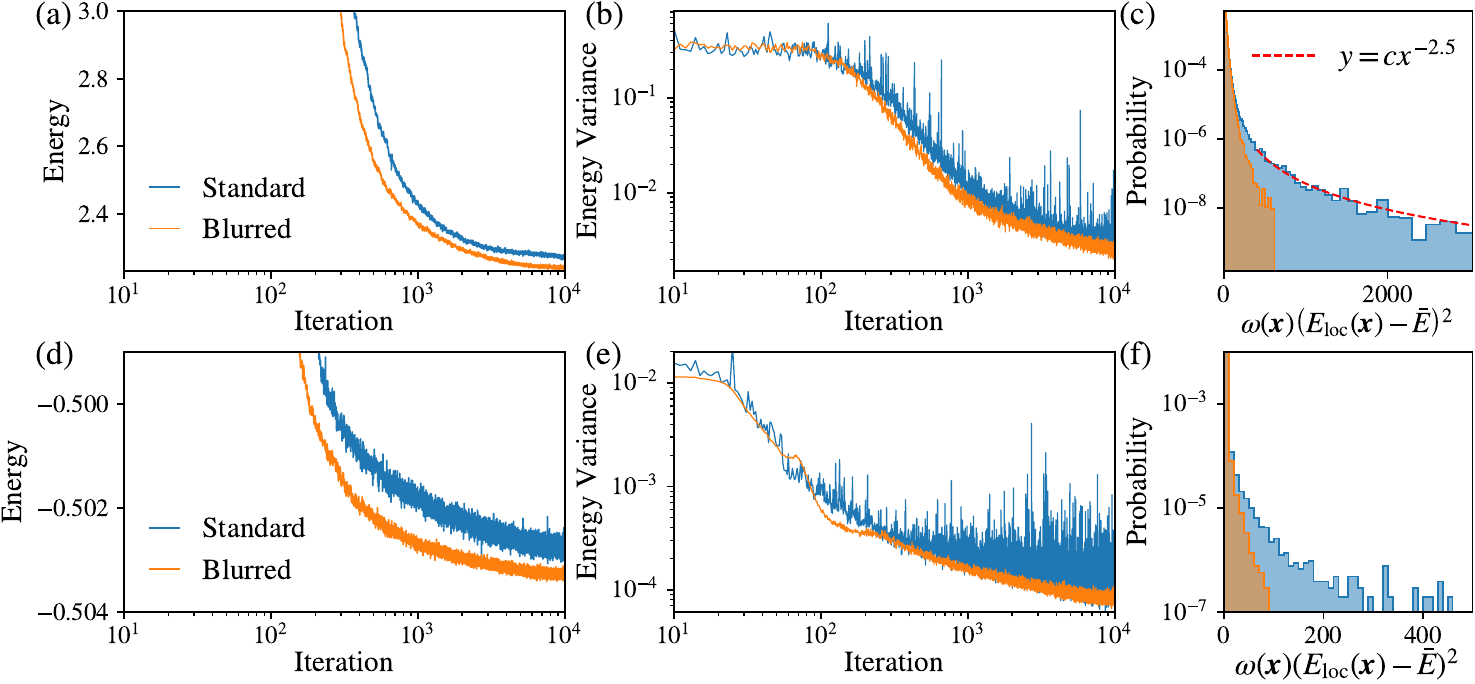}
    \caption{ground state optimization tests. 
    The top row shows a continuum example of the two-dimensional electron gas, with $N=45$, $r_s = 0.5$, and full spin polarization: (a) variational energy per electron, $E/N$, in Hartree, as a function of optimization iteration for standard and blurred sampling;
    (b) corresponding energy variance during optimization, illustrating the suppression of large stochastic spikes by blurred sampling; (c) the distributions of the energy-variance estimator $\omega(\config)\big(E_{\mathrm{loc}}(\config)-\bar{E}\big)^2$. 
    In (c), the red dashed line is a power-law fit of the state obtained by standard sampling, confirming a heavy-tail distribution with exponent $\alpha=3/2$, corresponding to a probability-density tail exponent of $5/2$.
    The bottom row shows a discrete example, the $J_1$-$J_2$ model on a $6\times 6$ lattice at the frustrated point $J_2/J_1=1/2$, with a similar study: (d) variational energy per site for optimization trajectories with standard sampling and blurred sampling;  
    (e) corresponding variance of the energy during optimization;  (f) distributions of the corresponding energy-variance estimators obtained at the end of the optimizations. 
    }
    \label{fig:fig_gs}
\end{figure*}

\subsection{Ground state optimization}
\label{ssec:gs-vmc}

\emph{Continuum-space example.}
Here we consider one of the iconic examples in the continuum, the two-dimensional electron gas, with Hamiltonian
\begin{equation}
\hat H
=
-\frac{1}{2}\sum_{i=1}^{N}\nabla_i^2
+\sum_{i<j} \frac{1}{|\vect{r}_i-\vect{r}_j|}
+ \text{b.g.},
\label{eq:2dheg_ham}
\end{equation}
in a periodic cell, where b.g.~denotes the background terms required to enforce charge neutrality. 
We use atomic units and follow conventional notations, with the electron density parametrized by
\(r_s = \frac{1}{\sqrt{\pi n}}\),
where \(n=N/A\) is the particle density in a square cell of area \(A\). 
We test a state-of-the-art neural quantum state, $(\mathrm{MP})^2\mathrm{NQS}$, as the trial wave function.
See, e.g., Ref.~\cite{ConorPRL2024}, for additional details. 
The blur strength $q$ and the blur distance $\epsilon$ are the two hyperparameters. We choose $q=0.5$ and $\epsilon=1$ here. The method remains effective even for small $q$ over a broad range of values.
A detailed analysis of the $q$ dependence of the estimator variance is given in App.~\ref{app:variance_variance}.

In Fig.~\ref{fig:fig_gs}(a), we show the optimization results for an $N=45$, $r_s=0.5$, spin-polarized system. We use the blur kernel defined in Eqs.~\eqref{eq:continuous_kernel} and \eqref{eq:blur_kernel}. 
It is clear from Fig.~\ref{fig:fig_gs}(a) that blurred sampling accelerates the optimization process compared with the standard method. 
To confirm that the slowing down is caused by nodal singularities, we also 
track the energy variance, which is crucial for reliable estimates of the ground-state energy and variance extrapolations.
As shown in Fig.~\ref{fig:fig_gs}(b), the energy-variance estimator under standard sampling exhibits repeated spikes, which disappear once blurred sampling is turned on.
Fig.~\ref{fig:fig_gs}(c) shows the single-sample distribution of
$\omega(\config)\big(E_{\mathrm{loc}}(\config)-\bar{E}\big)^2$,
where $\bar{E}$ is the energy estimate from the full sample and $\omega(\config)=1$ for standard sampling.
The expectation value of this quantity is the energy variance, with or without blurred sampling.
As discussed in Sec.~\ref{subsec:infinite_variance}, standard sampling is expected to produce a heavy tail with exponent
$\alpha=3/2$, which is confirmed by the fit shown as the dashed red line in Fig.~\ref{fig:fig_gs}(c).
With blurred sampling, the tail is strongly suppressed, the variance scaling is modified, and
the CLT is restored, effectively removing the infinite-variance problem caused by nodal pathologies.

\emph{Discrete-space example.}
For a discrete-space benchmark of ground state optimization, we consider the frustrated $J_1$--$J_2$ Heisenberg model,
with Hamiltonian 
\eq{
\hat H
=
J_1 \sum_{\langle i,j\rangle} \hat{\mathbf S}_i \cdot \hat{\mathbf S}_j
+
J_2 \sum_{\langle\!\langle i,j\rangle\!\rangle} \hat{\mathbf S}_i \cdot \hat{\mathbf S}_j ,
}{eq:j1j2_ham}
where $\langle i,j\rangle$ denotes nearest-neighbor pairs and $\langle\!\langle i,j\rangle\!\rangle$ denotes next-nearest-neighbor pairs. 
In our simulations, we study a $6\times 6$ square lattice with periodic boundary conditions.
We set $J_2/J_1 = 1/2$ and $J_1=1$, the most frustrated point, where the competition between $J_1$ and $J_2$ makes the optimization particularly challenging. 

In this case, although the problem may not exhibit strict support mismatch, frustration can still induce strong fluctuations in the wave function amplitude, making the estimators extremely noisy.
We therefore apply blurred sampling using a Hamiltonian-induced kernel. As in the continuum benchmark, the blurred estimator is implemented as a post-processing step that leaves the original sampling dynamics unchanged.
We took $q=0.5$ here, using 16384 samples. The optimizations use the SR method, with all hyperparameters kept identical between standard and blurred sampling. 
Details of the ansatz and optimization are provided in App.~\ref{app:ground_state_details}.
As in the continuum example, blurred sampling improves the stability and accuracy of the optimization, as shown in \Fig{fig:fig_gs}(d). We also monitor the energy variance, shown in \Fig{fig:fig_gs}(e). Large spikes appear in the standard method, whereas blurred sampling removes them. Finally, the distribution in \Fig{fig:fig_gs}(f) shows that the standard estimator has a heavy tail, which is suppressed by blurred sampling as expected.

The above two examples on ground-state VMC  
show that blurred sampling is a simple and effective stabilization method for SR optimization in both continuous and discrete configuration spaces. 
The statistical fluctuations originating from nodal singularities in ratio-type estimators are 
a very general feature, which can be remedied by the approach discussed here.
Because blurred sampling is implemented as a post-processing step with only modest overhead, it can be dropped into existing SR workflows without changing the sampler or variational ansatz. 
This suggests that blurred sampling can serve as a default in all such calculations in the future, whenever estimator noise or support-mismatch pathologies are a concern. 

\subsection{t-VMC: Previously Identified Difficult Examples}
In ground state optimization, nodal and support-mismatch pathologies primarily manifest as heavy-tailed fluctuations, noisy estimators, and slow convergence. In t-VMC, however, they can qualitatively alter the projected dynamics through biased force and QGT estimators, leading to incorrect time evolution even in the infinite-sample limit. 
Here, we first revisit two benchmarks where standard t-VMC is known to fail due to bias in the force estimators.

\begin{figure}[t]
    \centering    \includegraphics[width=1.0\columnwidth]{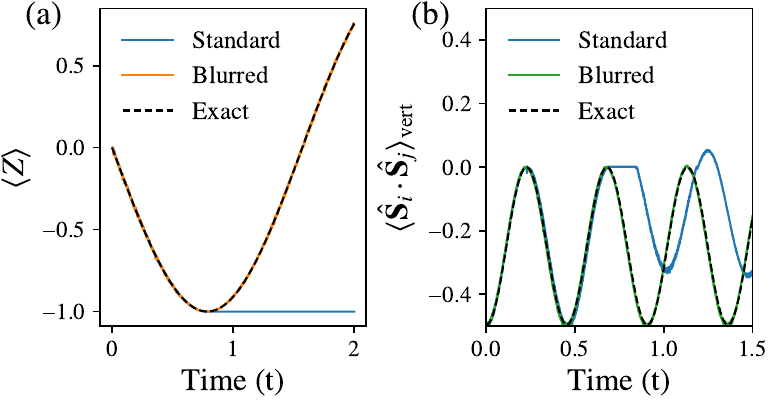}
    \caption{t-VMC results for two known benchmark problems. (a) We use the variational state $\ket{\ppsi} =\alpha\ket{0} + \beta\ket{1}$, with $(\alpha,\beta)=(1,1)$ at $t=0$ and evolve it under $U(t)=\exp\{-itY\}$ with t-VMC. As shown in Ref.~\cite{Filippo2023tVMC} the standard Monte Carlo estimator collapses due to finite-support bias as the variational state gets close to $\ket{0}$, causing the dynamics to diverge. With blurred sampling, the correct dynamics is recovered. 
    (b) A complex RBM with a single hidden unit fails to provide the correct dynamics under t-VMC for the $2\times2$ Heisenberg Hamiltonian~\cite{vrcan2025instability}, as the spin correlation diverges from the exact result using the standard sampler. The blurred sampling method produces the correct  dynamics. 
    In both cases, we use $q=0.5$ Hamiltonian-induced kernel for blurred sampling. 
   }
    \label{fig:Fig5}
\end{figure}

(i) \emph{Single spin.} Following Ref.~\cite{Filippo2023tVMC}, we evolve $\ket{\psi_\theta}=\alpha\ket{0}+\beta\ket{1}$ under $\hat H=Y$. 
As the state approaches $\ket{0}$, standard sampling collapses due to a support-mismatch bias. 
Blurred sampling removes the bias and reproduces the exact dynamics, as shown in \Fig{fig:Fig5}(a).

(ii) \emph{$2\times2$ Heisenberg quench.}
We next consider the anisotropic-bond Heisenberg model~\cite{vrcan2025instability},
\begin{equation}\label{eq:hm}
\hat H_\text{HM}
= \sum_{\langle i,j \rangle}
\hat{\mathbf S}_i \cdot \hat{\mathbf S}_j
+ \Delta J \sum_{(i,j)\in\mathrm{vert}}
\hat{\mathbf S}_i \cdot \hat{\mathbf S}_j ,
\end{equation}
where the second term runs over the vertical bonds of the lattice.
In Ref.~\cite{vrcan2025instability}, it was shown that a complex Restricted Boltzmann Machine (RBM) with a single hidden parameter fails to reproduce the correct real-time dynamics under t-VMC for the $2\times2$ system. 
The initial state is taken as the ground state of $\hat H$ at $\Delta J=0$ and is subsequently quenched to $\Delta J=-2$. 
The correlation $\sum_{(i,j)\in\mathrm{vert}}
    \langle\hat{\mathbf S}_i \cdot \hat{\mathbf S}_j\rangle \equiv \langle \hat{\mathbf S}_i \cdot \hat{\mathbf S}_j\rangle_{\mathrm{vert}}$ is measured as a function of time. 
Consistent with their findings, the standard Monte Carlo method breaks down due to support-mismatch bias, whereas blurred sampling successfully restores the correct dynamics, as shown in Fig.~\ref{fig:Fig5}(b).
(A more detailed analysis of this example is provided in App.~\ref{sec:breakdown} as a pedagogical example.)

\subsection{Solution of the Parity Mixing Problem}
We now turn to a larger-scale problem that captures a physically relevant mechanism: symmetry-sector mixing during non-equilibrium dynamics.
In many quantum quenches, the initial state that is being evolved may reside entirely within a specific symmetry sector, while the quench Hamiltonian couples different symmetry sectors to each other. 
The exact time evolution then necessarily involves coherent transfer of weight between them. 
Accurately describing such sector mixing is essential for capturing dynamical symmetry restoration, collective oscillations, and more generally symmetry-changing processes in non-equilibrium systems.

To construct a setting where this mechanism appears in a controlled manner, we consider an $n$-site, spin-$\frac{1}{2}$ system initialized in the equal superposition of all configurations $\config \in \{\pm1\}^n$ with even $Z$-parity,
\begin{align}
\ket{\psi^\text{even}_{\boldsymbol{\theta}}} 
= \sum_{\boldsymbol{\config} \in \mathcal X_{\mathrm{even}}} \ket{\boldsymbol{\config}},
\end{align}
where  $\mathcal X_{\mathrm{even}}=\{\config \in \{\pm1\}^n :\hat{\mathcal P}_Z \ket{\config}
=+\ket{\config}\}$ and the $Z$-parity operator is defined as $\hat{\mathcal P}_Z =\prod_{i=1}^{n} Z_i$.
Here $\ket{\config}$ denotes a computational-basis ($Z$-basis) state.

The system is evolved under the Hamiltonian of a transverse-field Ising model (TFIM) with periodic boundary conditions:
\eq{
\hat H_\text{TFIM} = J \sum_{i=1}^{n} Z_i Z_{i+1} + h \sum_{i=1}^{n} X_i.
}{}
Crucially, the spin-flip terms in $\hat H_{\mathrm{TFIM}}$ map even-parity configurations to odd-parity ones, so that $\hat H \ket{\psi^\text{even}_{\boldsymbol{\theta}}}$ has support on $
\mathcal X_{\mathrm{odd}}=\{\config \in \{\pm1\}^n :\hat{\mathcal P}_Z \ket{\config}
=-\ket{\config}\}$, while $\ket{\psi^\text{even}_{\boldsymbol{\theta}}}$ itself has support only on $\mathcal X_{\mathrm{even}}$.
As a consequence, the force estimator becomes systematically biased, since Monte Carlo sampling under $p(\config)$ cannot access configurations that carry nonzero weight in $\hat H_\text{TFIM}\ket{\psi^\text{even}_{\boldsymbol{\theta}}}$.

\begin{figure}[htb!]
    \centering
    \includegraphics[width=\linewidth]{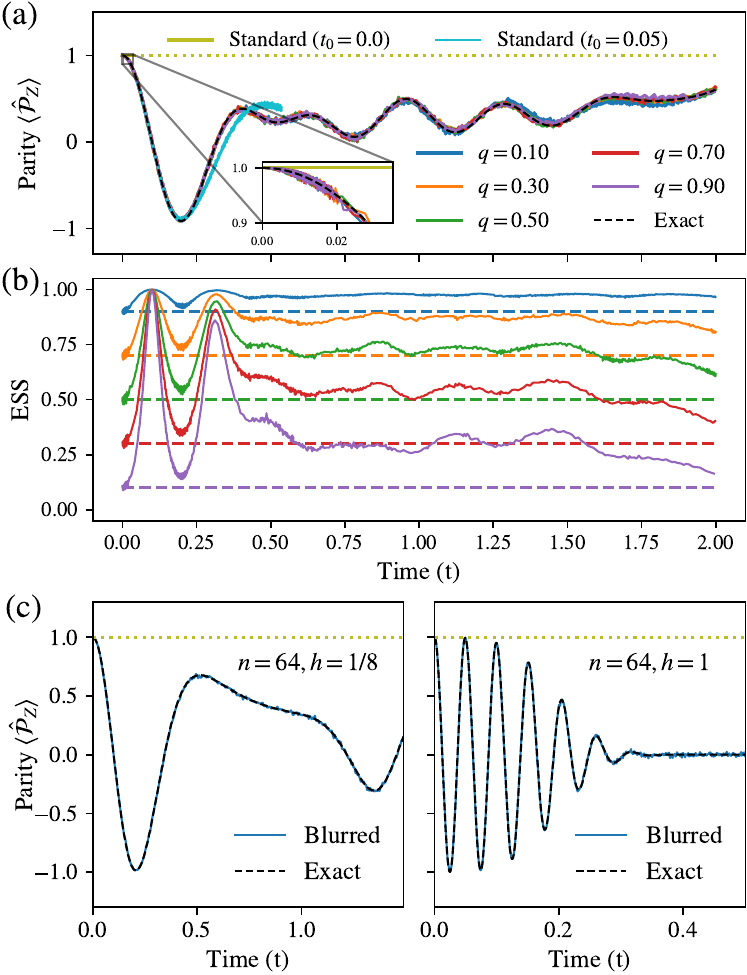}
    \caption{Accurate simulations of spin parity mixing with blurred sampling in t-VMC.
    (a) Time evolution of the parity operator for the quench of the $8$-spin state $\ket{\psi^\text{even}_{\boldsymbol{\theta}}}$ under the TFIM Hamiltonian at $J=h=1$. The variational state is an RBM with $
    32$ hidden units. Different lines indicate simulations with different values of the blur strength $q$. The inset shows that starting the t-VMC simulation with the standard estimator at $t_0=0$ leads to the immediate freezing of the dynamics due to the bias of the force. The dotted line shows the expected behavior of the parity for the standard $t=0$ dynamics. If we start the simulation at $t_0=0.05$, the standard estimator performs well for the initial ramp of the oscillation, but diverges for longer times. (b) The effective sample size $\mathrm{ESS} = \mathbb{E}[\omega(\config')]^2/\mathbb{E}[\omega(\config')^2]$. The dashed lines indicate the value $1-q$. At initialization, the $\mathrm{ESS}$ is exactly $1-q$. For all $q$, the $\mathrm{ESS}$ peaks whenever the parity goes through $0$, indicating that the state is balanced between odd and even parity sectors. (c) Time evolution of the parity operator using a Gaussian-state ansatz, with $n=64$. We show the slow ($h=1/8$) and fast ($h=1$) quench on the left and right, respectively. The dotted line indicates the expected freezing of the t-VMC dynamics with the standard estimator.
    Blurred sampling uses the Hamiltonian-induced kernel with $q=0.5$.
    }
    \label{fig:Fig6}
\end{figure}

In Fig.~\ref{fig:Fig6}, we investigate the  dynamics with t-VMC, by computing the time-dependent expectation value of the $Z$-parity under real-time evolution $U(t)=\exp(-it\hat H_{\mathrm{TFIM}})$, 
starting from $\ket{\psi^\text{even}_{\boldsymbol{\theta}}}$.
For the $n=8$ system, we employ a complex RBM as the variational ansatz. For this architecture, the initial state can be represented
analytically (see App.~\ref{app:rbm_initialization}) and the full dynamics can be simulated exactly with Majorana fermions (see App.~\ref{app:majorana}).
At $t=0$, the standard Monte Carlo sampling draws configurations only from the even-parity sector. 
Consequently, the parity remains constant, as evidenced by the flat curve in Fig.~\ref{fig:Fig6}(a), and the TDVP dynamics is, incorrectly, frozen. 
This is a support-mismatch problem; accordingly, simply making the sampler more overdispersed does not restore the missing odd-parity sector, as shown in App.~\ref{sec:overdispersed}.

One might expect that initializing the evolution slightly away from the pathological point, i.e., at a small $t_0>0$ where the state already contains a small odd-parity component, would restore the correct dynamics. 
However, we observe that standard sampling still fails: although odd-sector configurations are now formally present, their statistical weight remains extremely small, leading to a severely biased and noisy force estimator. 
The resulting trajectory deviates from the exact solution at later times as shown in Fig.~\ref{fig:Fig6}(a).

In contrast, blurred sampling restores access to the missing parity sector at every time step and faithfully reproduces the exact real-time dynamics over long times.
Moreover, the results are robust with respect to the blur strength $q$, which we vary in the range $q\in[0.1,0.9]$.
This demonstrates the stability of the method against changes in its only hyperparameter.
We also stress that all previous results presented in this work are similarly insensitive to the choice of $q$. 
This robustness stems from the bounded effective sample size and the structural elimination of estimator bias.
As shown in \Fig{fig:Fig6}(b), the effective sample size remains lower-bounded by $1-q$, in agreement with the theoretical expectation.
A more detailed analysis of the signal-to-noise ratio of the force estimator and the ESS as functions of $q$ is presented in App.~\ref{app:snr_force}.

To demonstrate scalability, we extend the system size to $n=64$ spins. 
For these larger systems, we find that the RBM ansatz is no longer sufficiently expressive to accurately capture the long time evolution.
To disentangle variational expressiveness from estimator bias in the t-VMC algorithm, we instead employ a Gaussian state ansatz tailored to the structure of the problem, which is capable of representing the exact state at all times (see App.~\ref{app:gaussian}).
As shown in Fig.~\ref{fig:Fig6}(c), blurred sampling accurately reproduces the parity-mixing dynamics for both a slow quench ($h=1/8$) and a fast quench ($h=1$), whereas the standard method fails in both cases.
This demonstrates that our method remains stable and accurate 
 at large system sizes and across different dynamical regimes.

\subsection{Unbiased Spin Relaxation Dynamics }\label{subsec:spin_relaxation}

We now consider a physically distinct regime in which the variational state is initially highly localized in configuration space. 
Such situations naturally arise in spin-relaxation dynamics, where the system starts from a nearly classical product state and evolves under a many-body Hamiltonian. 
Strong localization of the wave function also appears in neural quantum state simulations of quantum circuits, where one typically evolves a single computational basis state~\cite{medvidovic2021classical}. 
In all these cases, the state carries significant weight on only a small subset of configurations, and an accurate description of the time evolution requires efficient exploration of the surrounding states in the Hilbert space.
They thus provide realistic and stringent tests of the new method in a large class of important applications.

We also use this problem to provide an illustration of how the flexibility of the blurred sampling framework can be exploited to further enlarge the effective sampling support.
The Hamiltonian-induced blur ensures that $\mathrm{supp}(\hat H\ppsi)$ is sampled and therefore removes the bias in the force estimator. However, the QGT can still suffer from a support mismatch when $\text{supp}(\partial_{\theta_i}\ppsi(\config)\partial_{\theta_j}\ppsi(\config))\nsubseteq \supp(\hat H\ppsi)\cup \supp(\ppsi)$ (see further discussion in App.~\ref{app:qgt_blur}). Note that unlike the bias in the forces, the QGT pathology depends explicitly on the chosen parameterization of the state through the derivatives of the wave function.

We adopt a parameterization of our variational ansatz in which a symmetric complex RBM is multiplied with a factor $f_{\boldsymbol{\phi}}(\config)$, ${\boldsymbol{\phi}}\in\mathbb{C}^{n+1}$,
\begin{align*}
    \ppsi(\config) = f_{\boldsymbol{\phi}}(\config) \, \psi_\text{symm-RBM}(\config)\,.
\end{align*}
Two forms of $f_{\boldsymbol{\phi}}$ are considered, to help illustrate the effect of the randomized blurred sampling. In the first, different spin sectors are parameterized independently in $f_{\boldsymbol{\phi}}(\config)=\phi_{s(\config)}$, while in the second we break the independence, and let $f_{\boldsymbol{\phi}}=\phi_1 + \phi_3$ when $s(\config)=3$,
thereby
correlating $s=1$ to the $s=3$ spin sectors. (See App.~\ref{app:dressed_rbm} for further details.) Below we refer to the two parameterization schemes as \emph{uncorrelated} and \emph{correlated}, respectively.

The variational state is initialized in the fully polarized all-down product state, with exponentially small amplitudes assigned to all other configurations (see App.~\ref{app:dressed_rbm} for more details).
It is straightforward to see that,
under the TFIM Hamiltonian ($J$=$h$=$1$), the resulting state has  significant support mismatch between the QGT estimator and the forces estimator.

We exploit the flexibility of randomized blurred sampling to further enlarge the effective sampling support. 
We introduce $\lambda=\{s_1,\cdots, s_n\}, s_i=\pm 1$ as a flip mask, and denote $\lambda\cdot \config  = \{x_1s_1,\cdots, x_ns_n\}$ to be the configuration obtained by flipping the spins of $\config$ according to $\lambda$. 
The flip mask $\lambda$ is sampled uniformly in different spin sectors.
The parameterized blur kernel is defined as
\eq{
\begin{split}
K_{\lambda}(\config'|\config)=&(1-q_1-q_2) \delta_{\config',\config} + \frac{q_1}{N_\text{conn}} \sum_{y\in \mathcal{C}^\text{out}(\config)} \delta_{\config', y}\\
&+ q_2 \delta_{\config', \lambda\cdot \config},
\end{split}
}{}
where $q_1$ and $q_2$ control the strength of two distinct blurring mechanisms: 
$q_1$ introduces Hamiltonian-induced blurring as discussed in the previous section;
$q_2$ introduces a flip via the sampled mask $\lambda$. For each fixed $\lambda$, the support of $r_\lambda=\sum_\config K_\lambda(\config'|\config)p(\config)$ always contains $\mathrm{supp}(\hat H \ppsi)$, ensuring that the force estimator remains unbiased. The additional flip term effectively enlarges the accessible configuration space. 
Notice that  $q_2=0$ recovers the Hamiltonian-induced blurred sampling of the previous section.

\begin{figure}[t]
    \centering    \includegraphics[width=1.0\columnwidth]{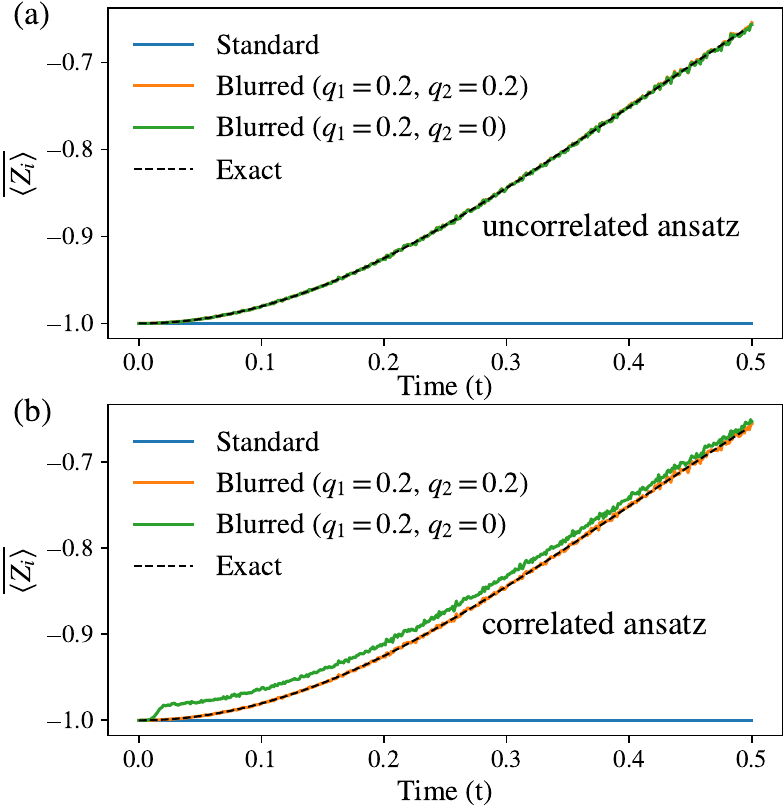}
    \caption{Solution of the spin relaxation problem. 
    The time evolution of 32 spins in the TFIM ($J=h=1$) is simulated,  initialized in the fully polarized all-down state. 
    The magnetization, defined as the site-averaged expectation value $\overline{\avg{Z_i}}$, is computed as a function of time, and compared against exact results obtained with Majorana fermions (see App.~\ref{app:exactZ}).
    Panels (a) and (b) show the results for two different parameterizations—respectively without and with correlations between parameters across different spin sectors. 
    For standard t-VMC simulations, the dynamics remain trapped in the initial state for both parameterizations, as expected (see also Ref.~\cite{Filippo2023tVMC}). In contrast, 
    blurred sampling with only the Hamiltonian-induced blur ($q_2=0$) performs well for the uncorrelated parameterization (panel a), but deviates from the exact solution in the correlated case (panel b). 
    Inclusion of randomized blurred sampling (with $q_1,q_2\neq 0$) successfully captures the relaxation dynamics in both cases.
   }
    \label{fig:Fig7}
\end{figure}

As shown in Fig.~\ref{fig:Fig7}(a,b), the standard estimator fails to reproduce the correct dynamics under $\hat H_\text{TFIM}$ and remains trapped at the initial state.  
Interestingly,  blurred sampling with only the Hamiltonian-induced term ($q_2$=$0$) already recovers the correct relaxation dynamics
in panel (a), with \emph{uncorrelated} parameterization.
This is because of the independent parameterization of the different spin sectors. Once a mild regularization is applied to the QGT, the remaining tangent-space directions are sufficiently well-conditioned to yield the correct parameter evolution, even without the additional randomized flips. 
In panel (b), with the \emph{correlated} parameterization, the Hamiltonian-induced blur alone ($q_2=0$) is no longer sufficient: the early-time dynamics becomes inaccurate due to the incomplete sampling of tangent-space directions. 
In this case, the inclusion of randomized blurred sampling restores the correct evolution. See App.~\ref{app:qgt_blur} for a detailed discussion of why randomized blurred sampling can yield the correct dynamics when the ansatz is sufficiently expressive, and App.~\ref{app:qgt_forces} for an analysis based on the QGT eigenvalues and variational forces.

\section{Discussion and concluding remarks}
In this work, we addressed one of the outstanding problems impeding a broad range of neural network variational calculations in quantum systems, including ground- and excited-states, finite-temperature properties, and imaginary- and real-time dynamics. We
introduced a flexible and general framework, blurred sampling, that regularizes both infinite-variance and
support-mismatch pathologies in variational Monte Carlo with a simple sampling scheme which is straightforward to implement.
By constructing an implicit reference distribution through a one-step local mixing of configurations, blurred sampling  overcomes the challenge of establishing an
explicit global reference measure and circumvents the competing
requirements that limit conventional importance-sampling approaches.

A key feature of blurred sampling is that it operates as a
post-processing step.
As a result, it preserves the original sampling dynamics and
autocorrelation structure, making it directly compatible with existing VMC workflows. This distinguishes blurred sampling from methods that rely on modified
sampling distributions or additional Markov chains, which can
significantly complicate both implementation and analysis.

The generality of the approach suggests several directions for further development.
While we focused on simple, physically motivated blur kernels, the
framework readily accommodates alternative choices, including
problem-specific kernels or adaptive constructions that optimize
statistical efficiency. 
For example, one could tune the blurring strength $q$ to target a fixed ESS. As long as the adaptive strategies depend only on information from previous steps, the method remains unbiased.

In the VMC setting, we show that blurred sampling addresses infinite-variance and support-mismatch pathologies, leading to stable optimization, a well-behaved energy-variance estimator, and reliable time evolution.
More generally, blurred sampling offers a promising route for reducing statistical errors in a broad range of observables and may be extended to more sophisticated VMC-based methods, such as projected t-VMC.
With the rapid growth of neural quantum state variational calculations, the potential gain in optimization effectiveness and efficiency can impact 
many applications across physics and chemistry.

Finally, although our analysis has been restricted to variational Monte Carlo, the underlying idea of implicit importance sampling through local
configuration mixing is applicable more broadly.
Potential extensions include projector quantum Monte Carlo methods,
finite-temperature simulations, and classical Monte Carlo problems
involving heavy-tailed observables.
Exploring these directions can provide a unified perspective on variance regularization across a wide range of stochastic simulation techniques. It is also of interest to examine the applicability of this approach to other optimization problems in machine learning and related areas of artificial intelligence.

\section*{Data Availability}
All numerical simulations on spin models were performed using NetKet~\cite{netket2:2019, netket3:2022}. 
Small-scale exact benchmarks were obtained with QuTiP~\cite{qutip5}, while larger-scale benchmarks were computed using the Majorana representation~\cite{han2024pfaffian, wan2025nishimori}. 
Further details of the numerical implementation are provided in Appendices, and the code required to reproduce these results is available at~\footnote{\url{https://github.com/therooler/nqs_blurred_sampling}}.

\begin{acknowledgments}
ZW is grateful for discussions with Miguel A. Morales, Conor Smith, Markus Holzmann and especially thanks Ao Chen for insightful discussions on the lattice model simulations. RW would like to thank Filippo Vicentini and Anka van der Walle for discussions. 
The Flatiron Institute is a division of the Simons Foundation. 
\end{acknowledgments}

\bibliography{ref}
\appendix

\clearpage

\onecolumngrid

\section{Detailed derivation and properties of blurred sampling} \label{app:derivation}
In this section, we derive the blurred sampling procedure and prove the properties stated in the main text. 
Blurred sampling does not modify the original sampler, which generates configurations $\config$ distributed according to $p(\config)$.
Throughout this section, we denote by $\config$ samples drawn from $p$, and by $\config'$ the configuration obtained after the post-processing step.

\subsection{Unit expectation of the reweighting factor}\label{app:unit}
We first verify that the reweighting factors have unit expectation, i.e., $\avg{\omega(\config')}_{r} = 1$, for the general blur kernel defined in \Eq{eq:general_kernel}.
Since $K(\config'|\config)$ is a Markov kernel, $\sum_{\config'} r(\config')=\sum_{\config'} \sum_{\config}K(\config',\config)p(\config)= \sum_\config p(\config)$, and hence $r$ is normalized whenever $p$ is normalized.
The reweighting factor is defined as $\omega(\config')=p(\config')/r(\config')$.
Therefore,
\begin{equation}
\avg{\omega(\config')}_r = \sum_{\config'} r(\config') p(\config')/r(\config')=\sum_{\config'}p(\config')=1.
\end{equation}

\subsection{Finite-sample unbiased estimators}\label{app:finite_sample_unbiased}
We first construct the finite-sample unbiased estimator for expectation values under the original distribution $p$.
Since $\avg{\omega}_r = 1$, we have
\eq{
\avg{A}_p = \frac{\avg{\omega A}_r}{\avg{\omega}_r}= \avg{\omega A}_r.
}{}
Therefore, the finite size estimator
\eq{
\widehat{\avg{A}_p} =  \frac{1}{N} \sum_{i=1}^{N} \omega(\config'_i) A(\config'_i)\equiv[\omega A]_r
}{}
is an unbiased estimator for $\avg{A}_p$ with $\mathbb E[{\widehat{\avg{A}_p}}]=\avg{A}_p$.

For covariance-type quantities $\avg{\avg{A,B}}_p\equiv \avg{AB}_p - \avg{A}_p\avg{B}_p$, we use the identities
$\avg{AB}_p = \avg{\omega AB}_r, \avg{A}_p = \avg{\omega A}_r, \avg{B}_p = \avg{\omega B}_r$. Hence,
\eq{
\avg{\avg{A,B}}_p= \avg{\omega AB}_r  - \avg{\omega A}_r\avg{\omega B}_r
}{}
We now use the identity $\avg{\omega}_r=1$ to rewrite the first term as
$\avg{\omega AB}_r = \avg{\omega}_r\,\avg{\omega AB}_r$, and add--subtract
$\avg{\omega^2 AB}_r$:
\eq{
\begin{split}
\avg{\avg{A,B}}_p&= \avg{\omega}_r\avg{\omega AB}_r - \avg{\omega^2 AB}_r \\
&-\avg{\omega A}_r\avg{\omega B}_r+\avg{\omega^2 AB}_r\\
&=\avg{\avg{\omega A, \omega B}}_r-\avg{\avg{\omega, \omega A B}}_r.
\end{split}
}{}
It is well known that the finite-sample unbiased estimator of the covariance $\avg{\avg{X_1, X_2}}_r$ is 
\eq{
\widehat{\avg{\avg{X_1, X_2}}}_r = \frac{N}{N-1}\left(\Navg{X_1 X_2}_r- \Navg{X_1}_r \,\Navg{X_2}_r\right).
}{}
Therefore we have that
\eq{
\begin{split}
\widehat{\avg{\avg{A,B}}}_p&=\widehat{\avg{\avg{\omega A, \omega B}}}_r-\widehat{\avg{\avg{\omega, \omega A B}}}_r\\
&=\frac{N}{N-1}\times\left(\Navg{\omega}_r\,\Navg{\omega AB}_r- \Navg{\omega A}_r\,\Navg{\omega B}_r\right)\\
&=\frac{N\,\Navg{\omega}_r^2}{N-1} \times
\left(\frac{\Navg{\omega A B}_r}{\Navg{\omega}_r} -\frac{\Navg{\omega A }_r}{\Navg{\omega}_r}\frac{\Navg{\omega B}_r}{\Navg{\omega}_r}\right),
\end{split}
}{eq:unbiased_cov}
is the finite-sample unbiased estimator for $\avg{\avg{A,B}}_p$.

In practice, one typically employs the self-normalized estimator (i.e., the right-hand side of the last line in \Eq{eq:unbiased_cov}), which is a biased estimator. 
However, this bias manifests itself as a common multiplicative factor affecting all covariance-like quantities.
Since the SR/TDVP update depends only on ratios of such quantities (e.g., through $\boldsymbol{S}^{-1}\boldsymbol{F}$), this common factor cancels in the parameter update.
Consequently, the resulting time evolution is unaffected by this overall factor.

\subsection{Bound of the reweighting factor and effective sample size}\label{app:bound}
Consider the distribution obtained after marginalizing the kernel $K$ in \Eq{eq:general_kernel}:
\begin{equation}
\begin{split}
    r(\config') &= \sum_{\config}K(\config'|\config)p(\config)\\
    &=\sum_{\config}\left((1-q)\,\delta_{\config',\config} p(\config) + q \, K_\mathrm{off} (\config'|\config) p(\config)\right)\\
    &= (1-q)p(\config') + q \, \sum_{\config}K_\mathrm{off} (\config'|\config) p(\config).
\end{split}
\end{equation}
Since $K_\mathrm{off} (\config'|\config) p(\config)\geq 0$,
we see that
\eq{
    r(\config') \geq (1-q)p(\config').
}{}
As a result, the reweighting factor is bounded by
\eq{
    \omega(\config')=\frac{p(\config')}{r(\config')}\leq\frac{p(\config')}{(1-q)p(\config')}=\frac{1}{1-q}.
}{}
Using the fact that  $p(\config')\geq 0$, we conclude that the reweighting factor is bounded as
\eq{
    0\leq  \omega(\config')\leq\frac{1}{1-q}.
}{}
Since the reweighting factors satisfy $\omega(\config) \in [0,\frac{1}{1-q}]$, we have the pointwise inequality
\eq{
    \omega(\config)^2 \leq \frac{1}{1-q}\omega(\config) .
}{}
Taking expectations with respect to $r$ yields
\begin{align}
\avg{\omega^2(\config')}_r \le \frac{1}{1-q}\avg{\omega(\config')}_r.
\end{align}
Using $\avg{\omega(\config')}_r=1$, we obtain
\begin{align}\label{eq:ess}
\mathrm{ESS}[\omega] = \avg{\omega(\config')}_r^2/\avg{\omega(\config')^2}_r\ge 1-q.
\end{align}

\subsection{Randomized blur construction}\label{app:randomized_blur}

For the randomized generalization of blurred sampling, we first recall the construction.
Let ${K_\lambda}$ be a family of blur kernels parameterized by $\lambda$, where each kernel satisfies the structure of \Eq{eq:general_kernel} and has sparse connectivity.
For each sample $\config_i \sim p(\config)$, we independently draw $\lambda_i \sim f(\lambda)$ from a normalized density $f$, and then generate $\config'_i \sim K_{\lambda_i}(\cdot|\config_i)$.
Let $r_\lambda(\config')=\sum_{\config}p(\config)K_\lambda(\config'|\config)$ denote the blurred distribution corresponding to a fixed $\lambda$.
We define the conditional reweighting factor
$\omega_\lambda(\config')=\frac{p(\config')}{r_\lambda(\config')}$.
Then the estimator $\Navg{\omega_\lambda A}_{f(\lambda)r_\lambda}$ remains finite-sample unbiased and equal to $\avg{A}_p$, 
\eq{
\avg{A(\config)}_p = \avg{\avg{\omega_\lambda(\config') A(\config')}_{r_\lambda}}_{f(\lambda)}=\avg{\omega_\lambda(\config') A(\config')}_{f(\lambda){r_\lambda}},
}{}
provided that no support-mismatch pathology occurs.
Similarly, the estimator in Eq.~\eqref{eq:unbiased_estimator} also applies under randomized blurred sampling, with $\Navg{\omega \cdot}_{r}$ replaced by $\Navg{\omega_\lambda\cdot}_{f(\lambda)r_\lambda(\config')}$.
As a direct application, the probabilistic blur construction
can be used to refine the continuum kernel introduced
in \Eq{eq:continuous_kernel},
where the displacement magnitude $\epsilon$ and coordinate directions $\mathbf e_{\alpha}$
are treated as random parameters $\lambda$.
By sampling these quantities from suitable distributions, the resulting blur approaches an effective Gaussian kernel.
In particular, this randomization restores rotational symmetry
at the level of the induced proposal distribution.

\subsection{QGT estimator from randomized blurred sampling}\label{app:qgt_blur}

Although the marginal distribution of $\config'$ is
$r(\config')=\sum_\lambda f(\lambda)\, r_\lambda(\config')$,
which coincides with that induced by the average kernel
$K_{\mathrm{mix}}(\config'|\config)=\sum_\lambda f(\lambda)\, K_\lambda(\config'|\config)$, the present construction differs in how the reweighting factor is evaluated.
In our scheme, the weight is evaluated conditionally on the realized value of $\lambda$, i.e., with respect to $r_\lambda(\config')$ rather than the marginal $r(\config')$.
As a result, the computational cost remains governed by the sparsity of each individual kernel $K_\lambda$, rather than by the mixed kernel $K_{\mathrm{mix}}$, which can be difficult to calculate or require an auxiliary Monte Carlo calculation.
This randomized formulation therefore represents a trade-off: while it does not fully expand the support at the level of individual blur kernels, it preserves sparsity and remains computationally feasible in regimes where the deterministic construction becomes costly.
As demonstrated in Sec.~\ref{subsec:spin_relaxation}, randomized blurred sampling successfully resolves cases in which Hamiltonian-induced blurred sampling fails to cure the QGT instability.

While the randomized formulation preserves sparsity and improves robustness in practice, its impact on the QGT estimator can be characterized more precisely.
In particular, although the marginal distribution $r(\config')$ may explore the full Hilbert space, the reweighting factor is evaluated conditionally on the realized value of $\lambda$, namely with respect to $r_\lambda(\config')$.
This conditioning leads to a modified structure of the QGT estimator, which we now make explicit.
For convenience, we denote by $\supp_\lambda$ the support of the conditional distribution $r_\lambda(\config')$.
We further introduce the indicator function
\eq{
\mathcal{I}_\lambda(\config) =
\begin{cases}
1, & \text{if } \config \in \supp_\lambda, \\
0, & \text{otherwise}.
\end{cases}
}{}

In the infinite-sample limit, the QGT estimator obtained from blurred sampling with a fixed realization of $\lambda$ is given by (the sum is only on $\supp_\lambda$ as we only have samples there) 
\begin{equation}
\begin{split}
S_{ij}(\lambda)
&=
\frac{\sum_{\config\in\supp_\lambda} r_\lambda(\config)\frac{\partial_{\theta_i}\ppsistar(\config)\partial_{\theta_j}\ppsi(\config)}{r_\lambda(\config)}
}
{\sum_{\config\in\supp_\lambda}r_\lambda(\config)\frac{|\ppsi(\config)|^2}{r_\lambda(\config)}}
-
\frac{
\left[\sum_{\config\in\supp_\lambda} r_\lambda(\config)
\frac{\partial_{\theta_i}\ppsistar(\config)\ppsi(\config)}
{r_\lambda(\config)}\right]
\left[\sum_{\config\in\supp_\lambda} r_\lambda(\config)
\frac{\ppsistar(\config)\partial_{\theta_j}\ppsi(\config)}
{r_\lambda(\config)}\right]
}{
\left[\sum_{\config\in\supp_\lambda}r_\lambda(\config)\frac{|\ppsi(\config)|^2}{r_\lambda(\config)}\right]^2
}
\\[4pt]
&=
\frac{\sum_{\config\in\supp_\lambda}
\braket{\partial_{\theta_i}\ppsi|\config}\braket{\config|\partial_{\theta_j}\ppsi}}
{\braket{\ppsi|\ppsi}}
-\frac{\braket{\partial_{\theta_i}\ppsi|\ppsi}}{\braket{\ppsi|\ppsi}}
\frac{\braket{\ppsi|\partial_{\theta_j}\ppsi}}{\braket{\ppsi|\ppsi}}
\\[4pt]
&=
\frac{\sum_{\config}\mathcal{I}_\lambda(\config)\,
\braket{\partial_{\theta_i}\ppsi|\config}\braket{\config|\partial_{\theta_j}\ppsi}}
{\braket{\ppsi|\ppsi}}
-\frac{\braket{\partial_{\theta_i}\ppsi|\ppsi}}{\braket{\ppsi|\ppsi}}
\frac{\braket{\ppsi|\partial_{\theta_j}\ppsi}}{\braket{\ppsi|\ppsi}},
\end{split}
\end{equation}
where we use the fact that $\text{supp}(\ppsi)\subseteq\text{supp}_\lambda$ for any $q<1$ since the blur kernel retains the original configuration with nonzero probability, which implies that $\metric=1$
for all $\config\in\supp(\ppsi)$. Consequently, the second term is unaffected by the blur and remains independent of \(\lambda\). 
Averaging over \(\lambda\) therefore gives
\eq{
S_{ij}^\text{r-blur}
=\avg{S_{ij}(\lambda)}_\lambda
=\frac{\sum_{\config} \avg{\mathcal{I}_\lambda(\config)}_\lambda \braket{\partial_{\theta_i}\ppsi|\config}\braket{\config|\partial_{\theta_j}\ppsi}}{\braket{\ppsi|\ppsi}}-\frac{\braket{\partial_{\theta_i}\ppsi|\ppsi}}{\braket{\ppsi|\ppsi}}\frac{\braket{\ppsi|\partial_{\theta_j}\ppsi}}{\braket{\ppsi|\ppsi}}.
}{eq:r_blur}
Compared with the exact QGT in \Eq{eq:force_and_QGT}, the deviation is 
\eq{
\Delta S_{ij}= \frac{\sum_{\config} \avg{1-\mathcal{I}_\lambda(\config)}_\lambda \braket{\partial_{\theta_i}\ppsi|\config}\braket{\config|\partial_{\theta_j}\ppsi}}{\braket{\ppsi|\ppsi}}.
}{}
Therefore, the randomized blurred QGT differs from the exact one only through configurations for which
$\avg{\mathcal I_\lambda(\config)}_\lambda<1$.

Importantly, this deviation does not necessarily invalidate the resulting dynamics.
We now show that, provided the variational ansatz is sufficiently expressive, randomized blurred sampling yields the exact projected TDVP evolution, despite modifying the QGT.
The key observation is that the induced modification can be interpreted as a positive deformation of the Hilbert-space metric underlying the TDVP derivation.

Recall that the TDVP equation in \Eq{eq:tdvp_equation} arises from minimizing the Hilbert-space distance between the variational tangent vector and the exact time-evolution direction.
In the tangent space, these directions are given by
\eq{
\begin{split}
\ket{\dot\ppsi}_\mathrm{var} &= \left(1-\frac{\ket{\ppsi}\bra{\ppsi}}{\braket{\ppsi|\ppsi}}\right)\frac{\sum_i\ket{\partial_{\theta_i} \ppsi }\dot{\theta_i} }{\sqrt{\braket{\ppsi|\ppsi}}} \\
\ket{\dot\ppsi}_\mathrm{exact} &= \left(1-\frac{\ket{\ppsi}\bra{\ppsi}}{\braket{\ppsi|\ppsi}}\right)\frac{-\xi\hat H\ket{\ppsi } }{\sqrt{\braket{\ppsi|\ppsi}}}. \\
\end{split}
}{}
The TDVP equation follows from minimizing the distance between these two tangent vectors, \eq{\mathcal{L}=|| \ket{\dot\ppsi}_\mathrm{var}-\ket{\dot\ppsi}_\mathrm{exact}||,}{}
where the distance is defined through the squared Hilbert-space norm $||\cdot||\equiv \sum_\config |\cdot(\config)|^2$.

If we replace the standard squared Hilbert-space norm by the weighted squared norm
$||\cdot||'\equiv \sum_\config \avg{\mathcal{I}_\lambda(\config)}_\lambda|\cdot(\config)|^2$,
the resulting TDVP equation reproduces precisely the QGT obtained from randomized blurred sampling.
To see this, consider the corresponding distance
\begin{equation}
\begin{split}
\mathcal{L}'&= \sum_{\config} \metric \left|\braket{\config|\dot\ppsi}_\mathrm{var}-\braket{\config|\dot\ppsi}_\mathrm{exact}\right|^2\\
&=\sum_{\config} \metric \left|\xi \left(\frac{\braket{\config|\hat H |\ppsi}}{\sqrt{\braket{\ppsi|\ppsi}}}-\frac{\braket{x|\ppsi}\braket{\ppsi|\hat H|\ppsi}}{\braket{\ppsi|\ppsi}^{3/2}}\right) +\sum_i \left(\frac{\braket{\config|\partial_{\theta_i}\ppsi}}{\sqrt{\braket{\ppsi|\ppsi}}}-\frac{\braket{x|\ppsi}\braket{\ppsi|\partial_{\theta_i}\ppsi}}{\braket{\ppsi|\ppsi}^{3/2}}\right) \dot\theta_i\right|^2
\end{split}
\end{equation}
Let $A_i(\config) \equiv \frac{\braket{\config|\partial_{\theta_i}\ppsi}}{\sqrt{\braket{\ppsi|\ppsi}}}-\frac{\braket{x|\ppsi}\braket{\ppsi|\partial_{\theta_i}\ppsi}}{\braket{\ppsi|\ppsi}^{3/2}}$, $B(\config)\equiv \left(\frac{\braket{\config|\hat H |\ppsi}}{\sqrt{\braket{\ppsi|\ppsi}}}-\frac{\braket{x|\ppsi}\braket{\ppsi|\hat H|\ppsi}}{\braket{\ppsi|\ppsi}^{3/2}}\right)$, we have 
\begin{equation}\mathcal{L}' = \vect{\dot\theta}^\dagger \vect{S}'\vect{\dot\theta} + \xi\vect{\dot\theta}^\dagger \vect{F}'+ \xi^*\vect{F}'^\dagger\vect{\dot \theta}+ \text{const.},
\end{equation} 
where $S'_{ij} = \sum_\config \metric A_i^*(\config) A_j(\config) $ and $F'_{i} = \sum_\config \metric A_i^*(\config) B(\config) $. 
Clearly, minimization of $\mathcal{L}'$ leads to the TDVP equation with $\vect{S}'$ as QGT and $\vect{F}'$ as variational force.
Therefore, we have 
\begin{align}
S'_{ij}
&=
\sum_{\config} \metric A_i^*(\config) A_j(\config)
\\
&=
\frac{\sum_{\config} \metric
\braket{\partial_{\theta_i}\ppsi|\config}
\braket{\config|\partial_{\theta_j}\ppsi}}
{\braket{\ppsi|\ppsi}}
-\frac{\braket{\ppsi|\partial_{\theta_j}\ppsi}}
{\braket{\ppsi|\ppsi}^{2}}
\sum_{\config} \metric
\braket{\partial_{\theta_i}\ppsi|\config}
\braket{\config|\ppsi}
\nonumber\\
&\quad
-\frac{\braket{\partial_{\theta_i}\ppsi|\ppsi}}
{\braket{\ppsi|\ppsi}^{2}}
\sum_{\config} \metric
\braket{\ppsi|\config}
\braket{\config|\partial_{\theta_j}\ppsi}
+\frac{\braket{\partial_{\theta_i}\ppsi|\ppsi}
\braket{\ppsi|\partial_{\theta_j}\ppsi}}
{\braket{\ppsi|\ppsi}^{2}}
\sum_{\config} \metric
|\braket{\config|\ppsi}|^2.
\end{align}
Using the fact that $\metric = 1$ for all $\config\in \supp(\ppsi)$,
the factors $\metric$ do not affect any term containing
$\braket{\config|\ppsi}$.
The last three terms reduce to their standard Hilbert-space expressions, and $S_{ij}'$ coincides exactly with the randomized blurred QGT
\(S^{\text{r-blur}}_{ij}\) in Eq.~(\ref{eq:r_blur}).

Randomized blurred sampling can therefore be viewed as a deformation of the Hilbert-space metric underlying the TDVP. By construction, $\avg{\mathcal{I}_\lambda(\config)}_\lambda>0$ for all $\config$, so the weighted norm remains strictly positive definite. 
Consequently,
\eq{
|| \ket{\dot\ppsi}_\mathrm{var}-\ket{\dot\ppsi}_\mathrm{exact}||=0 \,\Leftrightarrow\, || \ket{\dot\ppsi}_\mathrm{var}-\ket{\dot\ppsi}_\mathrm{exact}||'=0.
}{}
Therefore, if the variational ansatz is sufficiently expressive—namely, if there exists a parameter velocity $\dot{\boldsymbol{\theta}}^*$ such that $|| \ket{\dot\ppsi}_\mathrm{var}-\ket{\dot\ppsi}_\mathrm{exact}||=0$, then the same $\dot{\boldsymbol{\theta}}^*$ also solves the TDVP equation defined with the modified norm, $|| \ket{\dot\ppsi}_\mathrm{var}-\ket{\dot\ppsi}_\mathrm{exact}||'=0$ and vice versa.
Hence, randomized blurred sampling yields the correct parameter dynamics whenever the variational ansatz is sufficiently expressive.

In contrast, under support mismatch the standard VMC QGT corresponds to a deformation of the Hilbert-space metric with vanishing weights on configurations that are not sampled. The resulting metric therefore develops null directions and becomes degenerate, leading to potentially spurious dynamics even when the variational ansatz is sufficiently expressive.
Randomized blurred sampling removes this degeneracy. Although the resulting QGT differs from the exact one, the induced metric remains strictly positive definite and spans the correct tangent directions. In this sense, randomized blurred sampling acts as a natural regularization of the QGT, restoring stability while preserving the correct projected dynamics. 
A demonstration in the spin relaxation problem is presented in App.~\ref{app:qgt_forces}.

\clearpage
\section{Supplementary Results}

In this section, we present additional numerical results supporting the main text. We examine several estimator-level diagnostics, including the variance, signal-to-noise ratio, and effective sample size, to illustrate how blurred sampling addresses nodal and support-mismatch pathologies and remains robust over a broad range of $q$.
We also compare blurred sampling with overdispersed sampling, showing that the latter does not resolve the support-mismatch pathology and can suffer from an exponentially small effective sample size. Finally, we provide additional diagnostics for representative dynamical failures, which help identify cases where biased force or QGT estimates become important.

\subsection{Variance and ESS diagnostics for electron-gas system}\label{app:variance_variance}
\begin{figure}[h]
    \centering
    \includegraphics[width=0.95\linewidth]{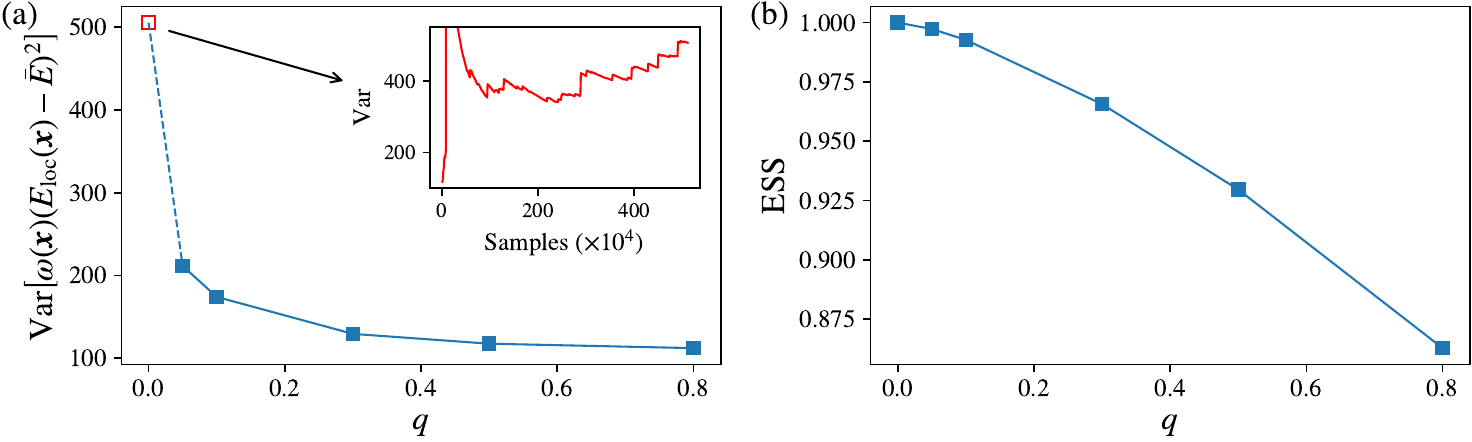}
    \caption{(a) Variance of the energy-variance estimator as a function of the blur strength $q$. Once $q$ is turned on, the variance of $\omega(\config)\left(E_{\mathrm{loc}}(\config)-\bar{E}\right)^2$ drops rapidly. The variance is estimated using $512\times 10^4$ samples. At $q=0$, the variance is expected to diverge theoretically. The inset shows that the estimated variance increases with the number of samples, and the large but finite value reported here reflects the finite sample size.  (b) Effective sample size as a function of $q$. The ESS decreases as $q$ increases. In these results, we use the same state as in \Fig{fig:fig_gs}(c) and $\epsilon=r_s$. 
    }\label{fig:app_var_heg}
\end{figure}

We quantify the effect of the blur kernel by varying the blur strength $q$ and measuring the variance of the energy-variance estimator in the electron-gas system. As shown in \Fig{fig:app_var_heg}, the variance is large at $q=0$, where standard sampling suffers from the nodal singularity, but drops sharply once blurred sampling is introduced. This is consistent with the theoretical expectation that the standard estimator has divergent variance; the large but finite value observed in \Fig{fig:app_var_heg} is a finite-sample remnant of this divergence. Any nonzero blur regularizes the singularity and restores ordinary central-limit-theorem scaling. The variance then remains stable over a broad range of $q$, indicating that the method does not require fine tuning of the blur strength. Over the same range, the ESS remains high, showing that the variance reduction is not offset by large reweighting fluctuations.

\vspace{-2em}
\subsection{Signal-to-noise and ESS diagnostics for parity-mixing problem}\label{app:snr_force}
\begin{figure*}[t]
    \includegraphics[width=0.95\linewidth]{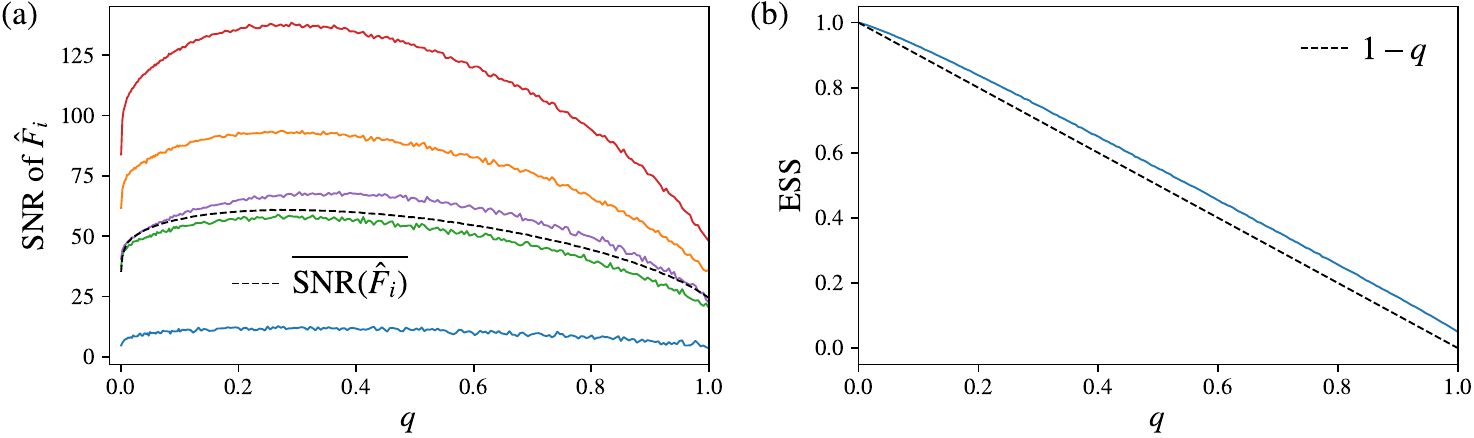}
    \caption{
    (a) The SNR of variational forces $\hat {\vect{F}}$ at time $t=0.2$ for the spin-parity quench in Fig.~\ref{fig:Fig6}, shown for different values of $q$. We show 5 representative components of the force vector as well as the average over all 592 parameters of the model. The SNR is significantly amplified when $q$ is turned on and remains stable over a broad range of $q\in[0.1,0.5]$. (b) The effective sample size of $\omega(\config)$ as a function of $q$, which is always larger than $1-q$ (the dashed line), as bounded by Eq.~\eqref{eq:ess}. Both the SNR of $\widehat{F}_i$ and the ESS are estimated from $N_s=262144$ samples for each $q$.
    }
    \label{fig:Fig6_snr}
\end{figure*}

As an additional diagnostic for the parity-mixing problem in \Fig{fig:Fig6}, we examine how the signal-to-noise ratio (SNR) of the force estimator depends on the blur strength $q$. 
We focus on the variational state at $t=0.2$, where the parity is $\hat{\mathcal{P}}_Z\approx -1$ and support-mismatch effects are expected to be pronounced. 
For each force component, we estimate
$$\mathrm{SNR}(\widehat{F}_i)=\frac{\langle\hat{F}_i(\config)\rangle_r^2}{\langle\hat{F}_i(\config)^2\rangle_r - \langle\hat{F}_i(\config)\rangle^2_r},
$$
as a function of $q$ with $\hat F_i(\config)\equiv \omega(\config) (\mathcal{O}^*_i(\config)-\bar{\mathcal{O}}^*_i)(E_\text{loc}(\config)-\bar{E})$.
As shown in \Fig{fig:Fig6_snr}(a), the SNR is strongly suppressed at $q=0$, reflecting the pathological variance of the standard estimator. 
With blurred sampling, the SNR increases rapidly and remains stable over an intermediate range of $q$, approximately between $q=0.1$ and $q=0.5$. 
For larger $q$, the SNR decreases again because the importance weights fluctuate more strongly, consistent with the reduction of the ESS shown in \Fig{fig:Fig6_snr}(b). 
Nevertheless, for all values of $q$, the ESS remains lower-bounded by $1-q$, in agreement with Eq.~\eqref{eq:ess}.
These results suggest that adaptively choosing $q$ to target an O(1) ESS could be a promising strategy.

\vspace{-1em}

\subsection{Comparison with overdispersed sampling}\label{sec:overdispersed}

\begin{figure}[b]
    \centering
    \includegraphics[width=0.9\linewidth]{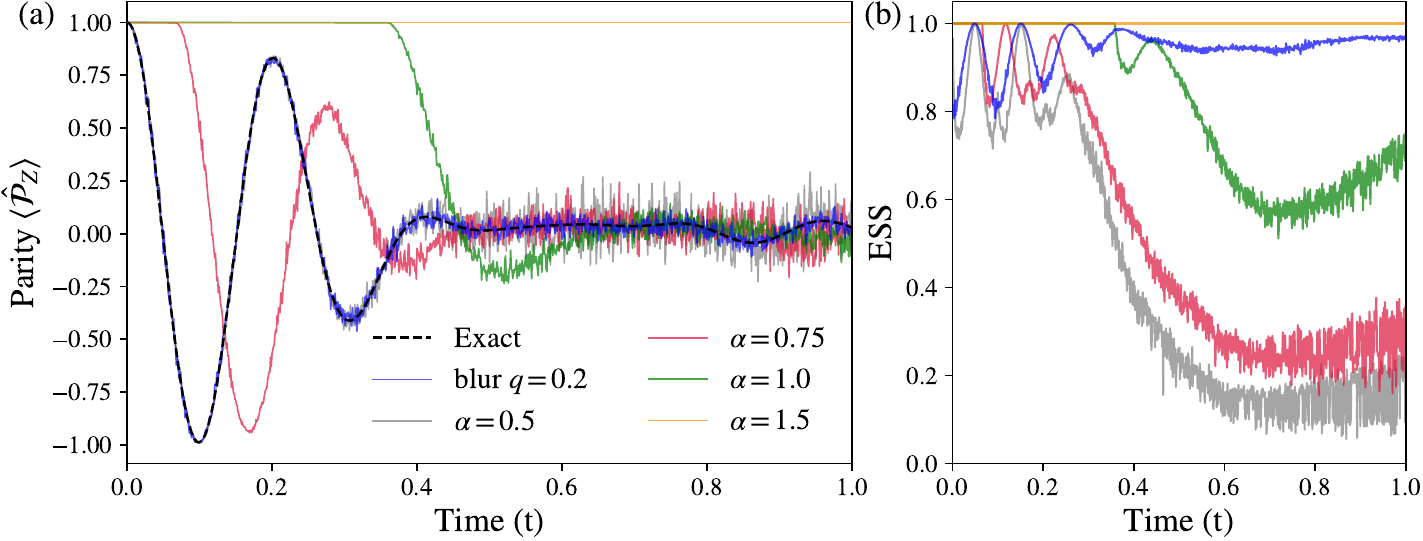}
    \caption{Comparison of overdispersed and blurred sampling in the parity-mixing problem. (a) The $Z$-parity dynamics simulated using the overdispersed sampling method with $\alpha=0.5,0.75,1.0,1.5$ and the blurred sampling method with $q=0.2$. The dashed line shows the exact solution. With $\alpha>0.5$, the overdispersed sampling results remain frozen at the initial value, whereas the blurred sampling follows the exact solution. (b) ESS for the different methods during time evolution. The overdispersed sampling method exhibits low ESS at late times, whereas the blurred sampling method remains above the bound $1-q=0.8$. The simulations are performed for a 16-spin system with $h=1$ and using the Gaussian ansatz. We use 1024 samples and Heun integrator with $\delta t = 10^{-3}$. }
    \label{fig:overdispersed}
\end{figure}

Here, we investigate the effectiveness of overdispersed sampling method~\cite{InuiPRRalphasampling,misery2025lookingelsewhereimprovingvariational}, a simple importance sampling method that replace the Born distribution by an overdispersed reference distribution, 
$r_\alpha(\config)
\propto
|\ppsi(\config)|^\alpha,
0\leq \alpha <2 $,
with the standard VMC method recovered at $\alpha=2$. 
For simplicity, we omit the normalization factors in the following discussion.
The corresponding reweighting factor is thus
$\omega_\alpha(\config)
=
p(\config)/r_\alpha(\config)
=
|\ppsi(\config)|^{2-\alpha}.$

Since $r_\alpha$ vanishes more slowly than $p(\config)=|\ppsi(\config)|^2$
near nodes, it samples configurations in the vicinity of nodal regions more frequently than conventional sampling. In this sense, overdispersed sampling is a direct importance-sampling approach: one modifies the Markov chain so that rare but statistically important configurations are visited more often, and then reweights observables back to the target Born distribution. Choosing $\alpha<2$ can reduce or remove the infinite-variance pathology by assigning more sampling weight to the nodal region.

However, this improvement comes with two important limitations. First, overdispersed sampling does not resolve support-mismatch bias in
discrete configuration spaces since the support of the distribution is unchanged. Second, the reweighting factor is a global function of the wave-function
amplitude. 
Since $\log|\ppsi(\config)|$ is typically extensive in
system size, the fluctuations of
$
\log \omega_\alpha(\config)
=
(2-\alpha)\log|\ppsi(\config)|
$
also grow with system size. Consequently, the effective sample size $\mathrm{ESS}_\alpha
=
\frac{\avg{{\omega}_\alpha}_{r_\alpha}^2}
{\avg{{\omega}_\alpha^2}_{r_\alpha}}
$
can decrease exponentially with system size for fixed $\alpha<2$.
For independent subsystems, this decay is strictly exponential in system size.
This is the central tradeoff of overdispersed sampling: reducing nodal singularities requires moving $\alpha$ away from $2$, but doing so increases the fluctuations of the reweighting factor away from the nodes. 

We illustrate these problems in Fig.~\ref{fig:overdispersed} by repeating similar simulations for the parity-mixing problem as in Fig.~\ref{fig:Fig6}.
For overdispersed sampling with $\alpha = 1.5$, the dynamics is frozen because of the support-mismatch problem.
For $\alpha= 0.75,1.0$, the dynamics are frozen at early times and recover only after some time. This is because the initial wave function does not have exact zeros; since we always add small perturbations to the parameters, overdispersed sampling amplifies these components. Only when $\alpha$ is very small do the dynamics recover at the beginning. We also compare with blurred sampling with $q=0.2$, which clearly gives the correct dynamics. 
In \Fig{fig:overdispersed}(b), we show the ESS for different methods. As $\alpha$ decreases, the ESS becomes smaller at late times, while blurred sampling remains lower-bounded by $1-q$.

\vspace{-1em}
\subsection{Breakdown of the Heisenberg dynamics}\label{sec:breakdown}

\begin{figure}[htb!]
    \centering
    \includegraphics[width=0.9\linewidth]{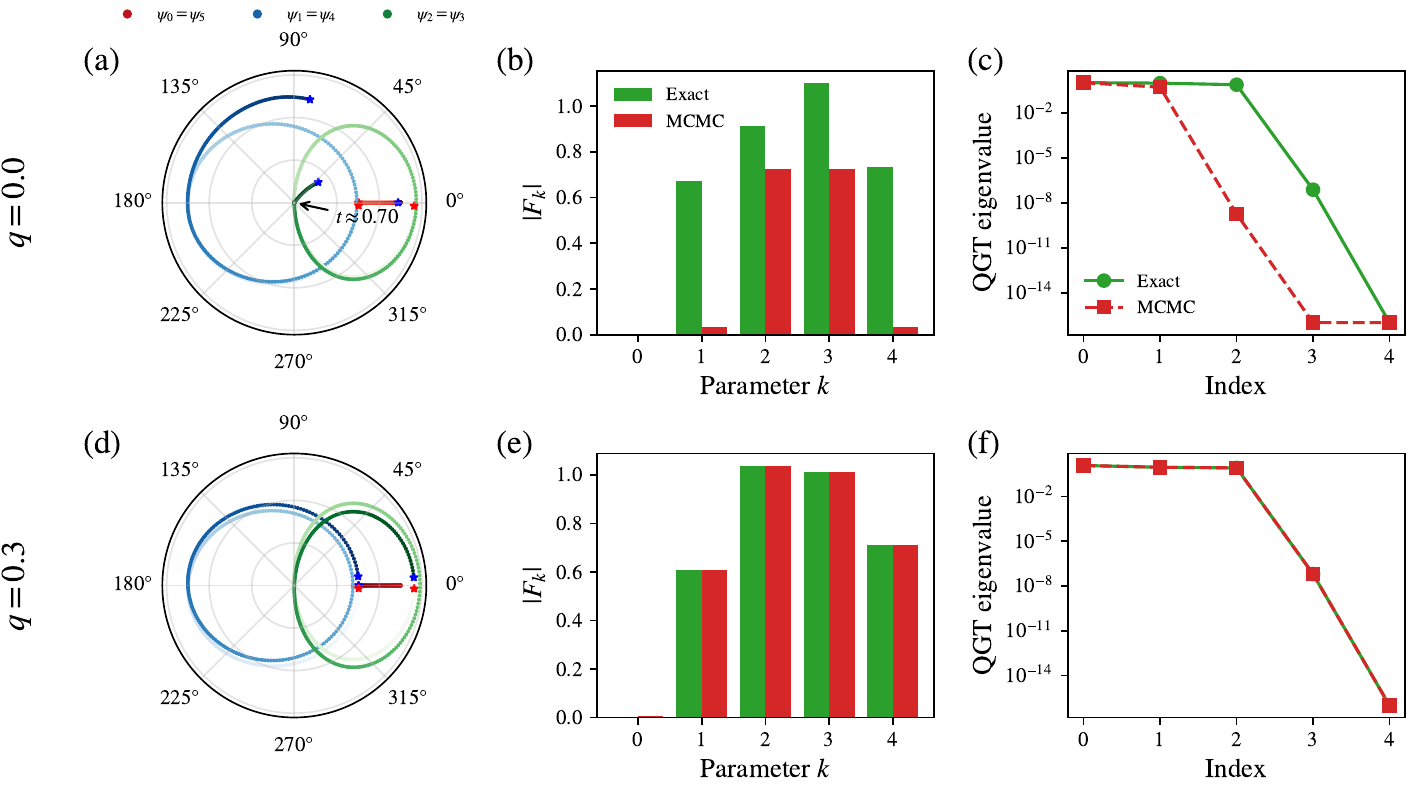}
    \caption{Diagnostics for the $2\times2$ Heisenberg quench at the breakdown time $t\approx0.70$, comparing standard tVMC ($q=0$, top row) with blurred sampling ($q=0.3$, bottom row). (a)(d) Trajectories of the three independent wave function
  amplitudes $\psi_0{=}\psi_5$, $\psi_1{=}\psi_4$, $\psi_2{=}\psi_3$ in the complex plane; red/blue stars mark the start and end of the dynamics, respectively. The dynamics repeatedly crosses the node, but around $t\approx0.70$ the standard sampler gets stuck at the node and the dynamics diverges.
  (b)(e) Magnitude of the force-vector components $|F_k|$,
  computed exactly by full enumeration of the Hilbert space (green) and estimated from MCMC samples (red).
  (c)(f) Corresponding eigenvalue spectra of the QGT, exact (green) vs. sampled (red).
}
    \label{fig:breakdown}
\end{figure}

Here we study the breakdown of the $2\times 2$ Heisenberg dynamics in \Fig{fig:Fig5}(b) near $t\approx 0.7$ in more detail. 
In \Fig{fig:breakdown}(a) and (d), we show the trajectories of the wave function amplitudes during the time evolution, after fixing the global phase by setting $\psi_0$ to be real and normalizing the wave function. At the breakdown point of $t\approx0.7$, parameter $\psi_2$ (which is equal to $\psi_3$ by symmetry) crosses zero in the complex plane, leading to a collapse of both the force and QGT estimators as shown in \Fig{fig:breakdown}(b)(c). 
Blurred sampling resolves this problem, as illustrated in \Fig{fig:breakdown}(e)(f).

\clearpage
\subsection{QGT bias and support of the forces in spin relaxation problem}\label{app:qgt_forces}

\begin{figure*}[htb!]
    \includegraphics[width=\textwidth]{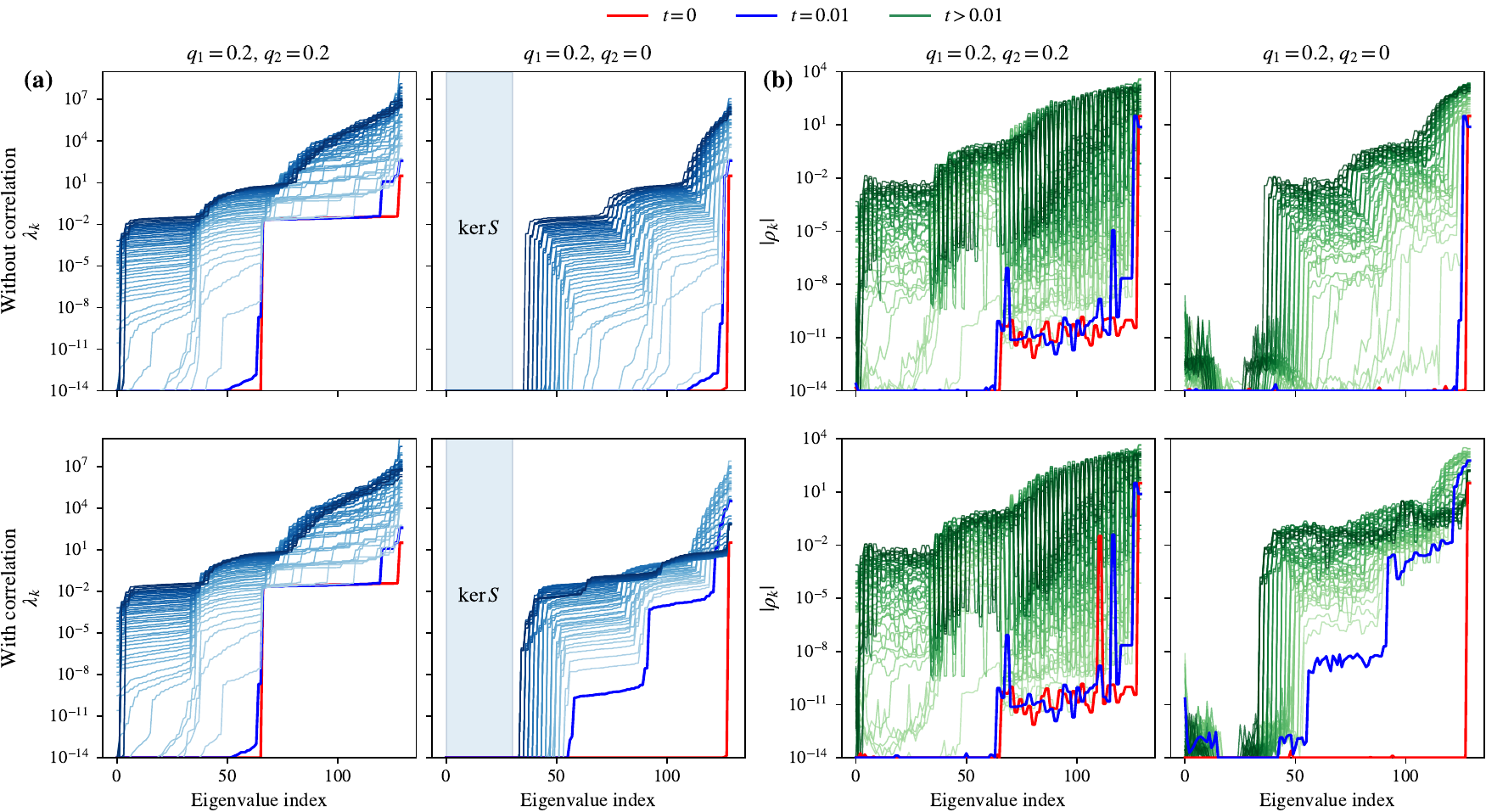}
    \caption{Comparison of the support of the QGT and the force vectors for the t-VMC dynamics of~\Fig{fig:Fig7} under blurred sampling. The first row corresponds to data for variational ansatz without correlation, whereas the second row corresponds to a variational ansatz with correlation between the spin sectors.
    (a) We diagonalize the QGT $\boldsymbol{S} = \boldsymbol{V}\boldsymbol{\Lambda}\boldsymbol{V}^{\dagger}$ and show the sorted eigenvalues $\lambda_k$ over the duration of the dynamics, with lighter colors indicating earlier times $t\leq 0.5$. We highlight the times $t=0$ and $t=0.01$ with red and blue lines, respectively. Note that for the Hamiltonian-induced blur ($q_2=0$, second column), the QGT has a large kernel for the total duration of the dynamics.
    (b) We transform the force vector to the eigenbasis of the QGT:
    $\boldsymbol{\rho} = \boldsymbol{V}^{\dagger} \boldsymbol{F}$ and show the components $\rho_k$ for the four cases studied here. The components are not sorted in magnitude, but correspond directly to the ordering of the eigenvalues in (a).}
    \label{fig:QGT_F_fig}
\end{figure*}

In this section, we argue why the Hamiltonian-induced blur sampling of~\Fig{fig:Fig7}(a) still produces the correct dynamics despite the QGT support mismatch problem, and why it fails in~\Fig{fig:Fig7}(b). In~\Fig{fig:QGT_F_fig} we calculate the eigenvalue spectrum of the QGT and investigate the support of the force vector in the corresponding eigenbasis. We see in panel (a) that without correlation in the ansatz (first row), the Hamiltonian-induced blur ($q_2=0$, second column) produces a QGT with extremely limited support, compared to the randomized blur ($q_2=0.2$, first column). This should result in a divergence in the dynamics from the correct result, but we see in~\Fig{fig:Fig7}(a) that this is not the case. The reason for this can be found in panel (b). We see that both blurred sampling schemes produce unbiased force vectors that are strongly peaked on a single mode for $t=0$ and $t=0.01$, with the support on other modes having $|\rho_k|<10^{-8}$. Due to this lack of support, the bias in the QGT does not affect the final force vector. However, when we look at the same quantities for the correlated ansatz (second row), a different picture emerges. For $t=0.01$, we see that the force vector has significant support on a large number of eigenmodes, which now get amplified incorrectly due to the QGT bias. The consequences are immediate: the dynamics in \Fig{fig:Fig7}(b) diverges around this point.

\clearpage

\clearpage
\section{Details of ground state optimization results}\label{app:ground_state_details}
\subsection{Numerical Details of the Two-Dimensional Electron Gas}\label{app:heg}

The wave-function ansatz used for the two-dimensional electron gas benchmark is the $(\mathrm{MP})^2\mathrm{NQS}$ ansatz~\cite{ConorPRL2024}, a neural backflow wave function. For a configuration
$\mathbf{R}=(\mathbf{r}_1,\mathbf{r}_2,\ldots,\mathbf{r}_N),$
the trial wave function is written as
\begin{equation}
\Psi(\mathbf{R})
=
\det\!\big[\phi_\alpha(\tilde{\mathbf{r}}_i)\big]_{\alpha,i=1}^N\,\times
\exp\!\left(\mathcal{U}_\text{NN}(\{\vect{\tilde{r}_i}\})-\sum_{i<j} u(r_{ij}) \right),
\label{eq:mp2nqs_ansatz}
\end{equation}
where $\phi_\alpha$ are the single-particle orbitals, $\tilde{\mathbf{r}}_i$ are the backflow-transformed coordinates,  $r_{ij}=|\mathbf{r}_i-\mathbf{r}_j|$ and $u(r)$ is the Jastrow factor, a symmetric two-body correlation function that captures short-range correlation effects and enforces the correct cusp structure in the wave function, which we parameterize using the form $
u(r)=\frac{2c_{\mathrm{cusp}} j_d^2\,e^{-j_b r}}{1+2 j_b j_d}\,\,\frac{1-e^{-r/j_d}}{r}$,
where $j_b$ and $j_d$ are variational parameters and $c_{\mathrm{cusp}}=1/3$ sets the electron-electron cusp strength~\cite{CCKjastrow}. 
The $\mathcal{U}_\text{NN}(\{\vect{\tilde r_i}\})$ is the neural-network Jastrow factor.
The single-particle orbitals $\phi_\alpha$ are expanded in a plane-wave basis as
\begin{equation}
\phi_\alpha(\mathbf{r})
=
\sum_{\mathbf{k}\in\mathcal{K}}
A_{\alpha,\mathbf{k}} \cos(\mathbf{k}\cdot\mathbf{r})
+
\sum_{\mathbf{k}\in\mathcal{K}}
B_{\alpha,\mathbf{k}} \sin(\mathbf{k}\cdot\mathbf{r}).
\end{equation}
Here $A_{\alpha,\mathbf k}$ and $B_{\alpha,\mathbf k}$ are real variational parameters. The wave vectors $\mathbf k$ are chosen from a truncated closed-shell momentum set $\mathcal K$, with $|\mathcal K|=97$, close to $2N$. 
Since the orbitals are represented in a real sine-cosine basis, the sum over $\mathbf k$ is taken over only one representative from each $\pm\mathbf k$ pair, avoiding double counting.

The backflow coordinates are defined by
$\tilde{\mathbf{r}}_i=\mathbf{r}_i+\boldsymbol{\xi}_i(\mathbf{R})$,
where the displacement field $\boldsymbol{\xi}_i(\mathbf{R})$ is produced by a message-passing neural network that depends symmetrically on the full electronic configuration. 
The neural network follows the same construction as Ref.~\cite{ConorPRL2024}, with one- and two-body streams $h_i$ and $h_{ij}$, periodic pair features. In our calculations, we use a setup with $d_1=16$ (one-body channels), $d_2=8$ (two-body channels), one message-passing layer, MLP width 16, attention width 16, and GELU activations. The neural Jastrow term uses one hidden layer with width 16.

In the simulations, we use 512 Monte Carlo samples per optimization step. The learning rate is initialized at $10^{-3}$ and decayed as $\text{lr}/(1+ \text{step}/1000)$. 
The wave function is optimized using SR method, with a diagonal regularization $\lambda {I} $ to the $S$-matrix where $\lambda=10^{-3}$. 
For Monte Carlo sampling, we use local Metropolis updates. Between two consecutive optimization steps, we perform 20 Metropolis steps to reduce autocorrelation. The proposal step size is adjusted adaptively during optimization to keep the harmonic average of acceptance rate close to 0.5.
The simulations with and without blurred sampling are initialized from the same state and use identical hyperparameters and sampling settings. The only difference is that, in the blurred-sampling runs, the blurred-sampling post-processing step is applied after every block of 20 local Metropolis updates.
The additional computational overhead of blurred sampling is modest. On 4 NVIDIA A100 GPUs, the standard calculation takes approximately $0.57$ s per optimization iteration, while the blurred-sampling calculation takes approximately $0.76$ s per iteration. This corresponds to a roughly $33\%$ increase in wall time, consistent with the expected constant-factor overhead from the post-processing step.

\subsection{Numerical Details of the $J_1$--$J_2$ Model}\label{app:J1J2}

We parametrize the variational wave function $\log\ppsi(\config)$ of a spin
configuration $\config$ on an $L\times L$ lattice using a patched Vision
Transformer architecture, following
Refs.~\cite{viteritti2023transformer,viteritti2025transformer}, together with
the factored attention mechanism introduced in
Refs.~\cite{rende2024mapping,rende2025queries}. The lattice, with $N=L^2$
sites, is partitioned into
\begin{equation}
  n = \frac{N}{b^2} = \left(\frac{L}{b}\right)^2
\end{equation}
patches of linear size $b$. Each flattened patch is linearly embedded into a
$d$-dimensional token $x_i\in\mathbb{R}^d$. The resulting sequence is processed
by $N_\ell$ transformer blocks of the form
\begin{equation}
  x_i' = x_i + \mathrm{Attn}\!\left(\mathrm{LN}(x)\right)_i,
  \qquad
  x_i'' = x_i' + \mathrm{FF}\!\left(\mathrm{LN}(x')\right)_i ,
\end{equation}
where $\mathrm{LN}$ denotes layer normalization. The feedforward block
$\mathrm{FF}$ is a two-layer MLP with intermediate width $4d$, mapping
$d\to 4d\to d$, and using a GELU nonlinearity.

In the factored-attention layer, the usual query-key mechanism is replaced by a
learned, translation-invariant mixing kernel. For each head
$a=1,\ldots,H$, the values
\begin{equation}
  v_i^{(a)} = W_v^{(a)} x_i
\end{equation}
are mixed according to
\begin{equation}
  o_i^{(a)}
  =
  \sum_{j=1}^{n}
  J^{(a)}_{i-j}\,v_j^{(a)} .
\end{equation}
Here $i-j$ denotes the relative displacement on the periodic patch lattice. We
parametrize the kernel as
\begin{equation}
  J^{(a)}_{\mathbf r}
  =
  p^{(a)}_{\mathbf r}\,
  \mathrm{softmax}_{\mathbf r}
  \!\left(-\alpha_a d_{\mathbf r}\right),
\end{equation}
where $p^{(a)}_{\mathbf r}$ is a learned per-head kernel amplitude,
$d_{\mathbf r}$ is the periodic Euclidean distance between patches, and
$\alpha_a$ is a learned per-head length-scale parameter initialized at
$\alpha_a=1$. This parametrization biases the initial kernel toward nearby
patches while retaining fully learnable translation-invariant weights. The
outputs from all heads are concatenated and projected back to dimension $d$.

After the final transformer block, the tokens are summed and normalized,
\begin{equation}
  z =
  \mathrm{LN}
  \!\left(
    \sum_{i=1}^{n} x_i^{(\mathrm{final})}
  \right),
\end{equation}
and two independent dense projections produce real and imaginary features,
\begin{equation}
  y_\mathrm{R} = \mathrm{LN}\!\left(W_\mathrm{R}z\right),
  \qquad
  y_\mathrm{I} = \mathrm{LN}\!\left(W_\mathrm{I}z\right).
\end{equation}
The complex log-amplitude is then given by
\begin{equation}
  \log\psi(\sigma)
  =
  \sum_k
  \log\cosh
  \!\left(
    y_{\mathrm R,k}
    + i\,y_{\mathrm I,k}
  \right).
\end{equation}

For the optimization shown in~\Fig{fig:fig_gs}(d), we use patch size $b=3$, $N_\ell=2$
transformer layers, $H=12$ factored-attention heads, and embedding dimension
$d=72$. For the system size considered here, this gives a variational ansatz
with $117048$ trainable parameters.

During optimization, we regularize the stochastic-reconfiguration linear system
by shifting the $S$-matrix estimator in Eq.~\eqref{eq:mc_estimator} by
$\lambda I$, with $\lambda=10^{-3}$. The learning rate is annealed with a
cosine-decay schedule from $\eta=10^{-2}$ to $\eta=10^{-3}$ over $10^4$
optimization steps.

\section{Details of the parity mixing problem in TFIM}\label{app:rbm}

We study the dynamics of the initial state chosen as the equal-weight superposition of computational-basis states with even $Z$-parity,
\eq{
\ket{\psi_0}
=\sum_{\config}
\frac{\hat{\mathcal{P}}_Z + 1}{2}
\ket{\config},
}{}
where $\hat{\mathcal{P}}_Z \equiv \prod_{i=1}^{N} Z_i$ denotes the $Z$-parity operator and
$\ket{\config}$ are computational-basis states with $x_i \in \{\pm 1\}$.
The time evolution is governed by the transverse-field Ising model (TFIM) Hamiltonian,
\eq{
\hat H_\text{TFIM} = J \sum_{i=1}^{N} Z_i Z_{i+1} + h \sum_{i=1}^{N} X_i,
}{eq:tfim}
where periodic boundary conditions are assumed.
In this setup, we are interested in how the $Z$-parity evolves under the dynamics, characterized by
\eq{
\mathcal{P}_Z(t) = \frac{\braket{\psi_0|e^{i \hat H_\text{TFIM}t}\hat{\mathcal{P}_Z}e^{-i \hat H_\text{TFIM}t}|\psi_0}}{\braket{\psi_0|\psi_0}}.
}{}

In the following, we describe the variational ansätze employed—including the RBM and Gaussian ansatz—and present the analytic solution.

\subsection{RBM ansatz and its initialization}\label{app:rbm_initialization}

We employ a restricted Boltzmann machine (RBM) with complex-valued parameters as the variational ansatz. The wave function is given by
\eq{
\psi_{\mathrm{RBM}}(\config) =
\exp\!\left(\sum_{i=1}^{N} a_i x_i\right)
\prod_{j=1}^{M}
2\cosh \left(b_j + \sum_{i=1}^{N} W_{ij}x_i\right).
}{eq:rbm_final}
where $a_i$, $b_j$, and $W_{ij}$ are complex parameters, and $M$ denotes the number of hidden units. We define the ansatz size as $\alpha\equiv M/N$.
The initialization is particularly simple. The action of the parity projector on a computational-basis state can be written as
\eq{
\frac{\hat{\mathcal{P}}_Z +1}{2} \ket{\config} = \cos\left(\frac{\pi}{4}(N-\sum_{i=1}^{N} x_i)\right)^2\ket{\config}.
}{}
Choosing the RBM parameters as
$a_i = 0$, $b_1=b_2 = \mi {\pi N}/{4}$,
$W_{i1} = W_{i2} = -\mi(\pi/4)$ (with all other parameters set to zero), the RBM wave function exactly reproduces these amplitudes, thereby initializing the variational state to the chosen parity sector.
For numerical stability, we add a small perturbation of order $10^{-3}$ to the initial parameters, which ensures that the logarithmic derivatives remain well defined.

\subsection{Exact solution via mapping to Majorana fermions}\label{app:majorana}
It is well known that the one-dimensional TFIM can be mapped to a system of free fermions. Here we follow the mapping of Ref.~\cite{wan2025nishimori} and directly introduce $2N$ Majorana fermions $\gamma_{2i},\gamma_{2i+1}$. 
The spin operators are represented as
\eq{
Z_{i+1} = (-\mathrm i)^i\gamma_0\gamma_1\gamma_2\cdots \gamma_{2i},\,\,
X_{i+1} = -\mathrm{i}\gamma_{2i}\gamma_{2i+1}.
}{}
One then obtains
\eq{
Z_i Z_{i+1} = 
\begin{cases}
-\mathrm{i} \, \gamma_{2i-1} \gamma_{2i}, & \text{for } i < N, \\
\mathrm{i} \, \hat{\mathcal{P}}_X \, \gamma_{2N-1} \gamma_0, & \text{for } i = N.
\end{cases}
}{}
where $\hat{\mathcal{P}}_X=\prod X_i$ is the conserved $X$-parity operator. For even system size $N$, the initial state $\ket{\psi_0}$ lies in the sector $\hat{\mathcal{P}}_X=1$.
Restricting to this sector, the Hamiltonian becomes
\eq{
\hat H_\text{TFIM} = -\mi J\sum_{i=0}^{N-2} \gamma_{2i+1} \gamma_{2i+2} + \mi J \gamma_{2N-1}\gamma_0 - \mi h \sum_i \gamma_{2i}\gamma_{2i+1}.
}{}
In this representation, the $Z$-parity operator reads 
\eq{
\mathcal{P} = (-\mi)^{N/2} \prod_{i=0}^{N/2-1}\gamma_{4i+1}\gamma_{4i+2}.
}{eq:parity_Z_majorana}
Although the Hamiltonian is quadratic in Majorana operators, so that the time-evolution operator is a fermionic Gaussian operator (a matchgate circuit), the initial state $\ket{\psi_0}$ is not Gaussian. We therefore exploit the symmetry of the $Z$-parity operator. 
Using the anticommutation relation $\{X_i,\hat{\mathcal{P}}_Z\}=0$, which implies 
\eq{
\hat{\mathcal{P}}_Z \hat H_\text{TFIM}(J,h) =\hat H_\text{TFIM}(J,-h)\hat{\mathcal{P}}_Z,
}{}
we obtain
\eq{
\begin{split}
\ket{\psi(t)} &\equiv e^{-i\hat H_\text{TFIM}(J,h) t}\ket{\psi_0}\\
&= e^{-i\hat H_\text{TFIM}(J,h) t}\frac{\hat{\mathcal{P}}_Z+1}{\sqrt{2}}\ket{+}\\
&=\frac{1}{\sqrt{2}}e^{-i\hat H_\text{TFIM}(J,h) t}\ket{+} + \frac{\hat{\mathcal{P}}_Z}{\sqrt{2}} e^{-i\hat H_\text{TFIM}(J,-h) t}\ket{+}.
\end{split}
}{}
Since the state $\ket{+}$ is Gaussian—being the ground state of the quadratic Hamiltonian $-\sum_i X_i=\sum_i \mi \gamma_{2i}\gamma_{2i+1}$. 
Its time evolution under the quadratic TFIM Hamiltonian remains Gaussian. 
We denote $e^{-i\hat H_\text{TFIM}(J,h) t}\ket{+}$ as $\ket{\psi_\text{G}^+(J,h, t)}$. 
The full wave function can then be written as
\eq{\ket{\psi(t)}=\frac{1}{\sqrt{2}}\ket{\psi_\text{G}^+(J,h, t)}+\frac{\hat{\mathcal{P}}_Z}{\sqrt{2}}\ket{\psi_\text{G}^+(J,-h, t)}.
}{}
The time-dependent $Z$-parity becomes:
\eq{
\begin{split}
\mathcal{P}_Z(t) 
&= \Re\braket{\psi_\text{G}^+(J,-h, t)|\psi_\text{G}^+(J,h, t)} \\
&+ \frac{1}{2}\braket{\psi_\text{G}^+(J,h, t)|\hat{\mathcal{P}}_Z |\psi_\text{G}^+(J,h, t)}\\
&+ \frac{1}{2}\braket{\psi_\text{G}^+(J,-h, t)|\hat{\mathcal{P}}_Z |\psi_\text{G}^+(J,-h, t)}.
\end{split}
}{}
The problem thus reduces to evaluating (i) the overlap between two Gaussian states and (ii) the expectation value of $\hat{\mathcal{P}}_Z$ in a Gaussian state. The overlap of two general fermionic Gaussian states (without $U(1)$ symmetry) can be computed using Pfaffian techniques; we refer the interested reader to Ref.~\cite{han2024pfaffian} for details.
The operator $\hat{\mathcal{P}}_Z$ is a product of Majorana operators [cf. \Eq{eq:parity_Z_majorana}], and its expectation value in a Gaussian state can be expressed as a Pfaffian of the corresponding covariance matrix $G_{i,j}\equiv\braket{\gamma_i\gamma_j}$ as
\eq{
\braket{\prod_i\gamma_{\alpha_i}}=\text{pf}\left[G_{\alpha_i,\alpha_j}\right].
}{}
Therefore, the time evolution of the $Z$-parity can be computed analytically for system sizes far beyond the reach of exact diagonalization.

\subsection{Gaussian ansatz and its initialization}\label{app:gaussian}
As shown in the main text, the RBM ansatz combined with blurred sampling performs well up to system sizes $N\sim 8$, where we employ a relatively large network with $\alpha=4$. 
For larger systems, however, the expressiveness of the ansatz becomes the primary bottleneck of the t-VMC approach.
To disentangle this limitation from other algorithmic aspects of t-VMC, we consider an alternative Gaussian ansatz. 
This ansatz permits efficient evaluation of wave function amplitudes and is capable of representing the exact time-evolved state $\ket{\psi(t)}$ at all times:
\eq{
\ket{\psi_\text{Gaussian} }=\ket{G^1}+\hat{\mathcal{P}}_Z\ket{G^2},
}{}
where $\ket{G^1}$ and $\ket{G^2}$ are Gaussian states parameterized as 
\eq{
\ket{G^{1,2}}= e^{\theta^{1,2}_{ij} \gamma_i\gamma_j}\ket{+},
}{}
with parameters $\theta^{1,2}$ being $2N\times 2N$ antisymmetric real matrices, each containing $N(2N-1)$ independent parameters.

The remaining task is to evaluate the amplitudes in the computational basis,
\begin{equation}\label{eq:gauss_ansatz}
    \braket{\config|\psi_\text{Gaussian}} =\braket{\config|G^1}+\hat{\mathcal{P}}_Z\braket{\config|G^2}.
\end{equation}
However, the basis state $\ket{\config}$ is not Gaussian.
Importantly, the wave function should preserve the $X$-parity, so that
$\braket{\config|\psi_\text{Gaussian}}$
should be the same as $\braket{\hat{\mathcal{P}}_X\config|\psi_\text{Gaussian}}$. 
Consequently, we may work in the symmetric combination 
\eq{\ket{\config}_+\equiv \frac{1}{\sqrt{2}}\ket{\config}+\frac{1}{\sqrt 2 }\ket{\hat{\mathcal{P}}_X\config}}{}
which is a Gaussian state. 
In fact, $\ket{\config}_+$ is the ground state of the quadratic Majorana Hamiltonian
\eq{
\mi \sum_{i=0}^{N-2} x_{i+1}x_{i+2}\gamma_{2i+1} \gamma_{2i+2} - \mi x_{N}x_{1} \gamma_{2N-1}\gamma_0.
}{}
This observation allows us to reformulate the amplitude evaluation entirely in terms of overlaps between Gaussian states, thereby enabling an efficient implementation within the t-VMC framework.
For numerical stability at initialization, we add a small perturbation of order $10^{-3}$ to the parameter $\theta^{1,2}_{ij}=0$ to ensure that the logarithmic derivatives remain well defined.

\section{Details of the spin relaxation problem in the TFIM}\label{app:spin_relaxation}
We study the real-time dynamics of the total $Z$ magnetization $\hat{\mathcal{Z}} =\sum_{i=0}^{N-1} Z_i$, starting from a computational-basis product state $\ket{\config_0}$ under the TFIM Hamiltonian.
The time-dependent expectation value is
\eq{
\mathcal{Z}(t)= \bra{\config_0}e^{\mi\hat H_\text{TFIM}(J,h)t}\hat{\mathcal{Z}}e^{-\mi\hat H_\text{TFIM}(J,h)t}\ket{\config_0}.
}{}
\subsection{Exact solution via mapping to Majorana fermions}\label{app:exactZ}
Using the anticommutation relation $\{X_i, Z_i\} = 0$, one obtains
$
Z_i e^{-i\hat H_{\mathrm{TFIM}}(J,h)t}
=
e^{-i\left(\hat H_{\mathrm{TFIM}}(J,h) - 2h X_i\right)t} Z_i,
$
which allows us to rewrite
\eq{
\begin{split}
\mathcal Z(t)
&= \sum_i
\bra{\config_0}
e^{i\hat H(J,h)t}
e^{-i\left(\hat H(J,h) - 2h X_i\right)t}
Z_i
\ket{\config_0}.
\end{split}
}{}
Since $\ket{\config_0}$ is an eigenstate of every $Z_i$ with eigenvalue 
$\config_0(i) \in \{\pm 1\}$, we obtain
\eq{
\begin{split}
\mathcal Z(t)
=
\sum_i \config_0(i)\,
\bra{\config_0}
e^{i\hat H(J,h)t}
e^{-i\left(\hat H(J,h) - 2h X_i\right)t}
\ket{\config_0}.
\end{split}
}{}
To proceed, we decompose the initial state into definite $X$-parity sectors,
$\ket{\config_0}=\frac{1}{\sqrt{2}}
\left(\ket{\config_0^+}+\ket{\config_0^-}
\right),
$
which yields
\eq{
\begin{split}
\mathcal Z(t)
=
\frac{1}{2}\sum_i &\config_0(i)
\Big(
\bra{\config_0^+}
e^{i\hat H(J,h)t}
e^{-i\left(\hat H(J,h) - 2h X_i\right)t}
\ket{\config_0^+}
\\
+
&\bra{\config_0^-}
e^{i\hat H(J,h)t}
e^{-i\left(\hat H(J,h) - 2h X_i\right)t}
\ket{\config_0^-}
\Big).
\end{split}
}{}
In this form, each contribution has a well-defined $X$-parity and can therefore be evaluated efficiently using the Majorana fermionic representation of the TFIM.

\subsection{Dressed RBM ansatz and its initialization}\label{app:dressed_rbm}
In the simulations, we employ a complex RBM with translation symmetry, denoted by $\psi_\text{symm-RBM}(\config)$ (see the definition in NetKet~\cite{netket3:2022}). 
For this symmetric RBM, we fix the visible bias parameter to zero and use a hidden-unit density $\alpha = 1$. 
When the RBM parameters are initialized close to zero ($\sim 10^{-3}$), the resulting wave function $\psi_{\text{symm-RBM}}(\config)$ approximates a uniform superposition over configurations. However, this ansatz alone cannot represent the desired initial state, which has support only on the fully polarized all-down configuration.

To incorporate this structure, we augment the symmetric RBM with a sector-dependent factor $f_{\boldsymbol{\phi}}(\config)$, parameterized by $n+1$ complex parameters $\boldsymbol{\phi} \in \mathbb{C}^{n+1}$:
\eq{
    \ppsi(\config) = f_{\boldsymbol{\phi}}(\config) \, \psi_\text{symm-RBM}(\config),\quad f_{\boldsymbol{\phi}}(\config)=\phi_{s(\config)}.
}{eq:dressed_rbm}
Here $s(\config) \equiv (\sum_i x_i + n)/2$ denotes the spin sector of configuration $\config$.
This construction assigns an independent complex amplitude to each spin sector, allowing the ansatz to selectively enhance or suppress entire sectors.
In particular, setting $\phi_0 = 1$ and $\phi_s = 0$ for $s \neq 0$ yields a wave function supported entirely on the fully polarized all-down configuration $\config = (-1, -1, \ldots, -1)$, corresponding to the $s=0$ sector.
In practice, to avoid exact zeros that may lead to numerical instabilities, we initialize $\phi_s = \epsilon = 10^{-5} \times 2^{-n/2}$ for $s \neq 0$, while fixing $\phi_0 = 1$ throughout the simulation.
Since $f_{\boldsymbol{\phi}}(\config)$ is independent across spin sectors in this parameterization, we refer to it as the \emph{uncorrelated} parameterization.

To further expose the bias of the quantum geometric tensor, we introduce correlations between different spin sectors in the parameterization.
In particular, we couple the single-spin-up sector to another sector, thereby breaking the independence of sector amplitudes. Concretely, we define
\begin{align*}
    f_{\boldsymbol{\phi}}(\config)=\begin{cases}
        \phi_1 + \phi_3 & \quad \text{if} \quad s(\config)=3\\
        \phi_{s(\config)} & \quad \text{otherwise},
    \end{cases}
\end{align*}
so that the amplitude of the $s=3$ sector explicitly depends on $\phi_1$, introducing correlations between the $s=1$ and $s=3$ sectors.
We refer to this as a \emph{correlated} parameterization.

Under the TFIM Hamiltonian, the resulting state has the following support structure:
\begin{align*}
    \supp(\ppsi)&\approx\{\config\in\{\pm 1\}^{n}: s(\config)=0\}\\
    \supp(\hat H_\text{TFIM}\ppsi)&\approx\{\config\in\{\pm 1\}^{n}: s(\config)\in \{0,1\}\}\\\supp(\partial_{\phi_i}\ppsi(\config)\partial_{\phi_j}\ppsi(\config))&=\{\config\in\{\pm 1\}^{n}\}.
\end{align*}
This reveals a pronounced support-mismatch problem both in variational force and QGT estimators.
In the following, we give further results to discuss the effectiveness of the blurred sampling method under both parameterizations.

\section{Numerical details of t-VMC simulations}\label{app:numerics}

The most important part of t-VMC involves the regularization of the $\boldsymbol{S}$-matrix in Eq.~\eqref{eq:tdvp_equation} so that it can be stably inverted. For this, we use the regularizer by Schmitt~\cite{schmitt2020quantum} with the hyperparameters $\mathrm{SNR}_{\mathrm{cutoff}}=2$, $\texttt{rcond}=10^{-14}$ and $\texttt{rcond\_smooth}=10^{-10}$ for every experiment reported in this work.
For sampling, we use Metropolis-Hastings MCMC with local updates depending on the Hamiltonian. In Table \ref{tab:numerics} we report the hyperparameters of the simulations. We emphasize that these are always kept exactly the same for the experiments with standard sampling and blurred sampling, hence any reported numerics differ only by the sampling scheme used. The code to reproduce all figures in this work can be found at~\cite{Note1}.
Finally, we note that the simulations reported in Figs.~2,3,5,6(a,b) and 7 were performed on a single NVIDIA RTX 6000 taking at most an hour of wall time. For Fig.~6(c) we used 4 NVIDIA H100s, with simulations lasting several hours at most.

\begin{table}[h]
    \centering
    \renewcommand{\arraystretch}{1.5}
    \begin{tabular}{|c|c|c|c|c|}
        \hline
        Figure & Ansatz $\ket{\ppsi}$  & MCMC & Integrator & Miscellaneous \\
        \hline\hline
        \Fig{fig:Fig5}(a) & $\alpha\ket{0} + \beta\ket{1}$ & $\begin{array}{c}
            \text{50\% single flip}\\
            \text{50\% double flip}\\
            N_s=8192
        \end{array}$& 
        $\begin{array}{c}
            \text{adaptive RK45} \\
            \texttt{rtol}=10^{-4}\\
            \min\delta t=10^{-5} \\
            \max\delta t=10^{-2}
        \end{array}$ & 
        \\ \hline
        \Fig{fig:Fig5}(b) & $\begin{array}{c}
            \text{RBM} \\
            \alpha = 1/4 \\
            \text{See Eq.~\eqref{eq:rbm_final}}
        \end{array}$ & $\begin{array}{c}
            \text{Two-site exchange}\\
            N_s=16384
        \end{array}$ & 
        $\begin{array}{c}
            \text{Heun} \\
            \delta t=10^{-3}
        \end{array}$ & 
        \\ \hline
        \Fig{fig:Fig6}(a,b) & $\begin{array}{c}
            \text{RBM} \\
            \alpha = 4 \\
            \text{See Eq.~\eqref{eq:rbm_final}}
        \end{array}$ & $\begin{array}{c}
            \text{50\% single flip}\\
            \text{50\% double flip}\\
            N_s=4096
        \end{array}$ & 
        $\begin{array}{c}
            \text{adaptive RK45} \\
            \texttt{rtol}=10^{-4}\\
            \min\delta t=10^{-5} \\
            \max\delta t=10^{-2}
        \end{array}$ & 
        \\ \hline
        $\begin{array}{c}
            \text{\Fig{fig:Fig6}(c)}  \\
            h=1/8
        \end{array}$& $\begin{array}{c}
            \text{Gaussian Ansatz} \\
            \text{See Eq.~\eqref{eq:gauss_ansatz}}
        \end{array}$ & $\begin{array}{c}
            \text{50\% single flip}\\
            \text{50\% double flip}\\
            N_s=8192
        \end{array}$ & 
        $\begin{array}{c}
            \text{Heun} \\
            \delta t=4\times10^{-3}
        \end{array}$ & $\begin{array}{c}
            \text{Used JAXMg~\cite{wiersema2026jaxmg} for distributed eigensolve.} \\
            \text{Sampled $|\ppsi|^2$ in 32-bit precision~\cite{solinas2026neural}.}
        \end{array}$
        \\ \hline
        $\begin{array}{c}
            \text{\Fig{fig:Fig6}(c)}  \\
            h=1
        \end{array}$& $\begin{array}{c}
            \text{Gaussian Ansatz} \\
            \text{See Eq.~\eqref{eq:gauss_ansatz}}
        \end{array}$ & $\begin{array}{c}
            \text{50\% single flip}\\
            \text{50\% double flip}\\
            N_s=8192
        \end{array}$ & 
        $\begin{array}{c}
            \text{Heun} \\
            \delta t=10^{-3}
        \end{array}$ & $\begin{array}{c}
            \text{Used JAXMg~\cite{wiersema2026jaxmg} for distributed eigensolve.} \\
            \text{Sampled $|\ppsi|^2$ in 32-bit precision~\cite{solinas2026neural}.}
        \end{array}$
        \\ \hline
 \Fig{fig:Fig7}(a,b)& $\begin{array}{c}\text{Dressed RBM} \\\alpha = 1\\
 \text{See Eq.~\eqref{eq:dressed_rbm}}
 \end{array}$& $\begin{array}{c}
            \text{50\% single flip}\\
            \text{50\% double flip}\\
            N_s=4096
        \end{array}$ & $\begin{array}{c}
            \text{Heun} \\
            \delta t=10^{-3}
        \end{array}$&\\\hline
    \end{tabular}
    \caption{Hyperparameters of t-VMC simulations}
    \label{tab:numerics}
\end{table}

\end{document}